\documentclass[journal]{IEEEtran}
\IEEEoverridecommandlockouts
\usepackage{lineno}
\usepackage{hyperref}
\usepackage{cite}
\usepackage{amsmath,amssymb,amsfonts,cases} 
\usepackage{amsmath}
\usepackage{amsthm}
\usepackage{subfig}

\usepackage{graphicx}
\usepackage{textcomp}
\usepackage{xcolor}
\usepackage{graphicx}
\usepackage{float}
\usepackage{amsmath,mleftright}
\usepackage{amsfonts,amssymb}
\usepackage{mathrsfs}
\usepackage{mathtools}
\usepackage{algorithm}
\usepackage{algorithmic}
\usepackage{bm}
\usepackage{multirow}
\usepackage{array}
\usepackage{amssymb}
\usepackage{amsmath}
\usepackage{cite}
\usepackage{url}
\usepackage{xcolor}
\usepackage{cite,graphicx,amsmath,amssymb}
\usepackage{fancyhdr}
\usepackage{mdwmath}
\usepackage{mdwtab}
\usepackage{caption}
\usepackage{amsthm}
\usepackage{setspace}
\usepackage{bm}
\usepackage{mathtools}
\usepackage{dsfont}
\usepackage{bbm}
\usepackage{framed}
\allowdisplaybreaks[4]
\newtheorem{remark}{Remark}
\newtheorem{theorem}{Theorem}

\newtheorem{lemma}{Lemma}

\newtheorem{corollary}{Corollary}

\makeatletter
\newcommand{\biggg}{\bBigg@{3}}
\newcommand{\Biggg}{\bBigg@{3.5}}
\makeatother
\makeatletter
\renewcommand{\maketag@@@}[1]{\hbox{\m@th\normalsize\normalfont#1}}%
\makeatother
\newcounter{problem}
\newcounter{save@equation}
\newcounter{save@problem}
\makeatletter
\newenvironment{problem}
{\setcounter{problem}{\value{save@problem}}%
  \setcounter{save@equation}{\value{equation}}%
  \let\c@equation\c@problem
  \subequations
}
{\endsubequations
  \setcounter{save@problem}{\value{equation}}%
  \setcounter{equation}{\value{save@equation}}%
}

\begin{document}
\title{Segmented Waveguide-Enabled Pinching-Antenna Systems (SWANs) for ISAC}
\author{Hao Jiang, Chongjun Ouyang, Zhaolin Wang, Yuanwei Liu,~\IEEEmembership{Fellow,~IEEE}, \\
Arumugam Nallanathan,~\IEEEmembership{Fellow,~IEEE}, Zhiguo Ding,~\IEEEmembership{Fellow,~IEEE}, and Robert Schober,~\IEEEmembership{Fellow,~IEEE}
\thanks{H. Jiang, C. Ouyang, and A. Nallanathan are with the School of Electronic Engineering and Computer Science, Queen Mary University of London, London, E1 4NS, U.K. (email: \{hao.jiang, c.ouyang, a.nallanathan\}@qmul.ac.uk).}
\thanks{Z. Wang and Y. Liu are with the Department of Electrical and Electronic Engineering, The University of Hong Kong, Hong Kong (email: \{zhaolin.wang, yuanwei\}@hku.hk).}
\thanks{Z. Ding is with the Department of Electronic and Electrical Engineering, The University of Manchester, M1 9BB Manchester, U.K. (e-mail: zhiguo.ding@manchester.ac.uk).}
\thanks{R. Schober is with the Institute for Digital Communications, Friedrich-Alexander-University Erlangen-Nurnberg (FAU), Germany (e-mail: robert.schober@fau.de).
} \vspace{-2.2em}
}

\maketitle
\begin{abstract}
A segmented waveguide-enabled pinching-antenna system (SWAN)-assisted integrated sensing and communications (ISAC) framework is proposed.
Unlike conventional pinching antenna systems (PASS), which use a single long waveguide, SWAN divides the waveguide into multiple short segments, each with a dedicated feed point.
Thanks to the segmented structure, SWAN enhances sensing performance by significantly simplifying the reception model and reducing the in-waveguide propagation loss.
To balance performance and complexity, three segment controlling protocols are proposed for the transceivers, namely i) \emph{segment selection} to select a single segment for signal transceiving, ii) \emph{segment aggregation} to aggregate signals from all segments using a single RF chain, and iii) \emph{segment multiplexing} to jointly process the signals from all segments using individual RF chains.
The theoretical sensing performance limit is first analyzed for different protocols, unveiling how the sensing performance gain of SWAN scales with the number of segments.
Based on this performance limit, the Pareto fronts of sensing and communication performance are characterized for the simple one-user one-target case, which is then extended to the general multi-user single-target case based on time-division multiple access (TDMA).
Numerical results are presented to verify the correctness of the derivations and the effectiveness of the proposed algorithms, which jointly confirm the advantages of SWAN-assisted ISAC.
\end{abstract}


\IEEEpeerreviewmaketitle
\section{Introduction}
To meet the demanding requirements of future communication networks, multiple-input multiple-output (MIMO) technologies play an indispensable role in the evolution of telecommunications \cite{dang2020should}.
The evolution of telecommunication technologies is characterized by advancements in MIMO technology, which have enhanced diversity gains, multiplexing gains, and array gains, forming a technical path from MIMO to massive MIMO and gigantic MIMO \cite{bjornson2025enabling}.

To harness the benefits of MIMO while minimizing implementation cost, NTT DOCOMO introduced pinching antenna systems (PASS) into the telecommunications research community in 2021 \cite{pinching_antenna1, ding2025flexible}.
Unlike conventional MIMO, PASS is endowed with channel reconfigurability, i.e., the ability to manipulate the wireless channel to a desirable condition.
In particular, PASS comprises dielectric waveguides for low-attenuation wired propagation and low-cost pinching particles for signal radiation.
From the perspective of signal propagation, signals first propagate through a low-attenuation waveguide that extends over tens of meters, and then are released into free space at the positions of the pinching particles, referred to as pinching antennas (PAs).
Due to the meter-scale length of the waveguides, large-scale fading can be mitigated.
Moreover, since waveguides and PAs are low-cost components, the implementation cost of PASS remains moderate.
With these advantages, PASS has emerged as a promising and practical reconfigurable antenna (RA) system for future communication networks \cite{liu2025pinching_tut}.

\subsection{Prior Works}
Building on the proven advantages of PASS, numerous studies have demonstrated its strong theoretical potential \cite{liu2025pinching_tut, xu2025generalized}.
More specifically, the authors of \cite{ding2025flexible} first studied PASS for wireless networks, confirming the benefits of PASS over conventional antenna systems. 
Subsequently, the authors of \cite{ding2025los} demonstrated that blockages, previously considered a negative factor in conventional MIMO scenarios, can be leveraged by PASS to mitigate multi-user interference, thereby increasing the performance gain over conventional MIMO.
This feature is also referred to as environment division multiple access (EDMA) \cite{ding2025environment}.
On top of the above, the author of \cite{ouyang2025array} offered answers to the fundamental question of PASS: how the array gain scales with the number of PAs and the spacing between PAs.
Moreover, joint optimization of digital beamforming and PA positioning, known as pinching beamforming, has drawn growing attention in recent studies.
For instance, a performance analysis framework was proposed in \cite{tyrovolas2025performance} to maintain the benefits of pinching beamforming.
Then, the authors of \cite{wang2025modeling} presented a modeling framework for in-waveguide propagation and beamforming design, while the authors of \cite{bereyhi2025mimo} proposed downlink and uplink beamforming designs for a multi-user MIMO scenario.
Furthermore, a generalized pinching beamforming framework based on element-wise iteration was proposed by \cite{sun2025multi}.
Moreover, it has been shown that the reconfigurability of PASS benefits a wide spectrum of applications, including physical layer security \cite{wang2025pinching}, machine learning \cite{guo2025graph}, etc.

Despite the extensive research on PASS for communications, its application to sensing remains an underexplored yet crucial research area, since sensing is emphasized in IMT-2030/6G \cite{kaushik2024toward}.
In particular, the authors of \cite{ding2025pinching} first derived the Cramér–Rao bound (CRB) for the localization error in PASS, which serves as a lower bound on the variance of the estimation error of any unbiased estimator.
This work revealed a fundamental tradeoff between the communication and sensing objectives in PASS. Specifically, the PA position that maximizes the communication performance, i.e., directly above the target, is not the same position that minimizes the sensing CRB.
Building on this work, the authors of \cite{jiang2025pinching} confirmed this finding from a Bayesian CRB perspective, which exploited prior knowledge on the target position distributions to relax the dependence of the CRB on the sensing parameters and obtained a general bound regardless of the unbiasedness of the estimator.
Furthermore, the authors of \cite{wang2025wireless} proposed a particle swarm optimization (PSO) algorithm to minimize the CRB, demonstrating that the CRB in PASS sensing is more robust to initial positioning errors than MIMO sensing.
In addition to the above, integrated sensing and communications (ISAC) paradigms have also drawn significant attention.
In particular, the authors of \cite{qin2025joint} and \cite{zhang2025integrated} first considered a downlink ISAC system in which pinching beamforming was leveraged to maximize throughput while ensuring sensing constraints.
Moreover, the authors of \cite{ouyang2025rate} depicted the rate region of PASS-assisted ISAC systems, where the inner and outer bounds were derived to underpin the sensing-throughput tradeoff.
Adopting CRB as a sensing metric, the authors of \cite{li2025pinching} presented an ISAC framework that employed a PASS transmitter and a uniform linear array (ULA) receiver.
Moreover, the authors of \cite{jiang2025pinching_tracking} explored the tracking functionality to dynamically locate a malicious user, thereby ensuring the covertness of transmissions.

\subsection{Motivations and Contributions}
However, existing research on PASS-assisted sensing was mainly based on an ideal uplink reception model, which ignores the inter-antenna radiation (IAR) effect \cite{ouyang2025uplink}, which occurs when multiple antennas are attached to one waveguide.
In particular, the received signal captured by one PA and guided via waveguides can be leaked/re-radiated by other PAs, making modeling signal reception more complicated and introducing inaccuracies in the signal models.
A potential approach to address this issue is the implementation of a single PA per waveguide \cite{ouyang2025rate}. 
However, this method restricts the subsequent use of multiple PAs for enhancing reception.

To address this issue, a segmented waveguide-enabled pinching antenna
system (SWAN) was proposed by the authors of \cite{ouyang2025uplink, xu2025generalized}, in which the original long waveguide guide is divided into multiple short segmented waveguides having their own feed points.
In this design, the segments are isolated to decouple in-waveguide propagation, enabling the use of multiple PAs, while avoiding complex electromagnetic modeling at each PA.
In addition to the enabling role that SWANs play in the uplink scenario, the short length of each segment also reduces the in-waveguide loss, especially for long waveguides.
Last but not least, in terms of reliability, replacing a faulty segment is far more cost-effective and less disruptive than replacing the entire long waveguide.
To leverage these advantages, this paper investigates the performance of SWANs for ISAC applications.
The contributions of this paper can be summarized as follows:
\begin{itemize}
    \item We propose a SWAN-assisted ISAC system serving multiple downlink users while sensing a target in monostatic fashion.
    Three operating protocols are proposed for the transceivers, namely i) \emph{segment selection (SS)} to select one segment for signal transceiving, ii) \emph{segment aggregation (SA)} to aggregate signals from all segments using a single RF chain, and iii) \emph{segment multiplexing (SM)} to jointly process signal from all segments using individual RF chains.
    
    \item We first characterize the theoretical sensing performance limits of the three protocols.
    The performance gains of SWAN over conventional PASS are derived for all three protocols, demonstrating how SWAN’s benefits scale with the number of segments.
    
    \item We then characterize Pareto fronts for the sensing and communication performance. 
    For the SS protocol, a closed-form solution is obtained. For SA, we derive closed-form PA placements on the echo-receiving side and solve the remaining variables element-wise. For SM, we derive the optimal receive and transmit beamformers via a subspace method and solve the PA placement problem element-wise.
    
    \item We further extend our study to a multi-user time-division multiple access (TDMA) system, introducing a pinching multiplexing scheme in which PA positions are jointly optimized across all time slots.
    The resulting scheduling problem is solved via an element-wise method under an optimal power allocation policy.
    
    \item Numerical results show that:
    i) SWAN outperforms the conventional PASS design by reducing in-waveguide loss, especially in large service area scenarios;
    ii) the sensing performance of the three protocols does not vary monotonically with the number of segments; and
    iii) due to the different degrees of freedom (DoF) available for optimization, SM achieves the best performance but at the highest computational cost, followed by SA with moderate complexity, and SS with the simplest optimization.
\end{itemize}

\subsection{Organization and Notations}
The remainder of this paper is organized as follows. 
Section \ref{sect:system_model} presents the system model for SWAN-assisted ISAC. 
Section \ref{sect:analystical} highlights the gain of SWAN from a performance analysis perspective. 
Section \ref{sect:two-user} depicts the Pareto front of sensing and communication in SWAN. 
Section \ref{sect:pinch_multiplexing} elaborates on the proposed pinch multiplexing method for the multi-user case. 
Numerical results are provided in Section \ref{sect:results}, and conclusions are drawn in Section \ref{sect:conclusions}.
\begin{figure*}
    \centering
    \includegraphics[width=0.6\linewidth]{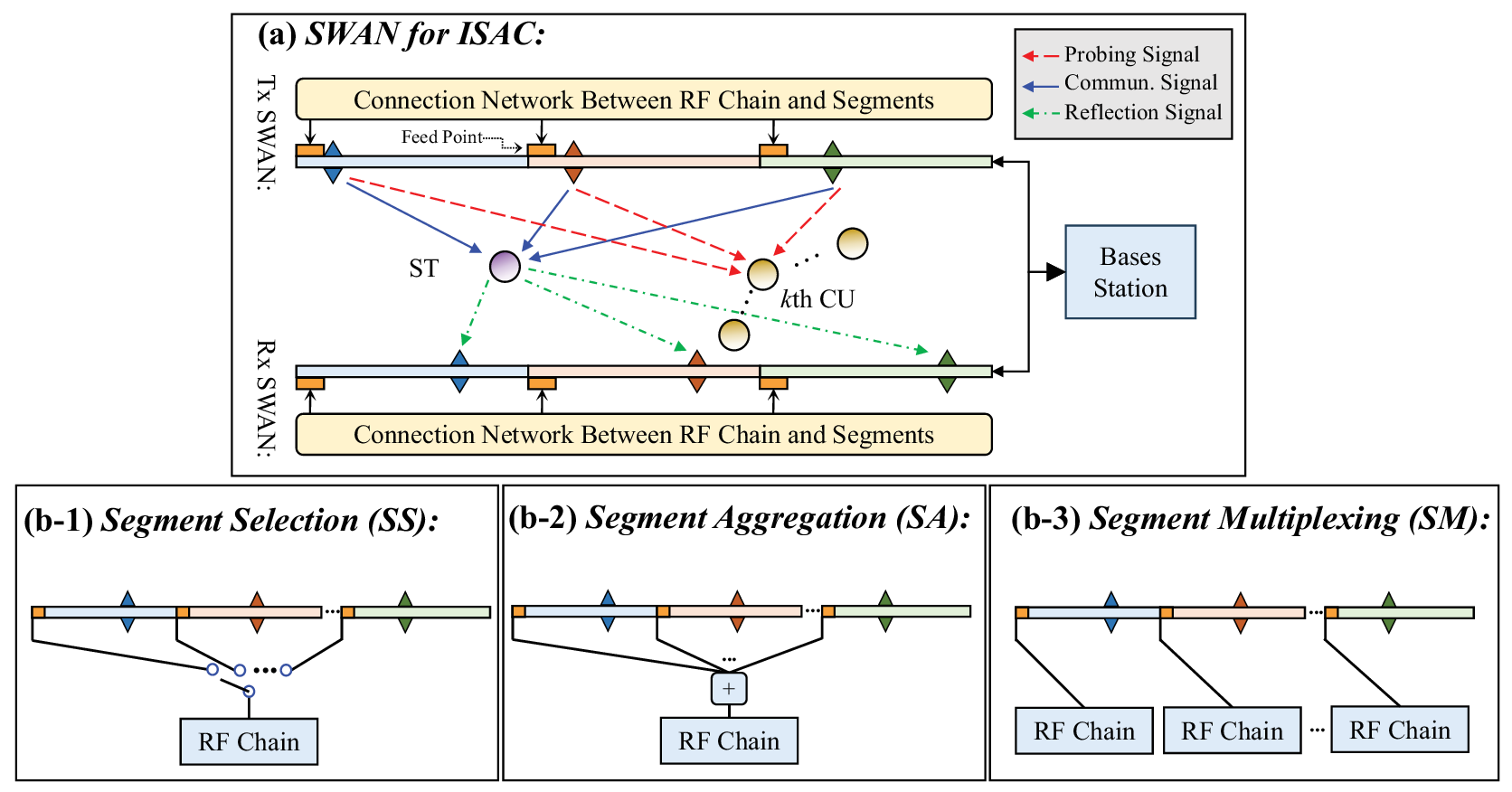}
    \caption{Illustration of the system model (see (a)) and the three protocols proposed for SWAN (see (b-1), (b-2), and (b-3)).}
    \label{fig:system_model_and_three_protocols}
    \vspace{-1.5em}
\end{figure*}

\textit{Notations:}
Scalars, vectors, and matrices are denoted by lower-case, bold-face lower-case, and bold-face upper-case letters, respectively.
$\mathbb{C}^{M \times N}$ and $\mathbb{R}^{M \times N}$ denote the space of $M \times N$ complex and real matrices, respectively.
$(\cdot)^\mathrm{T}$, $(\cdot)^*$, and $(\cdot)^\mathrm{H}$ denote the transpose, conjugate, and conjugate transpose, respectively.
$|\cdot|$ and $\lceil \cdot \rceil$ represent the absolute value and the ceiling function, respectively.
For a vector $\mathbf{a}$, $[\mathbf{a}]_i$ and $\left\| \mathbf{a} \right\| $ denote the $i$-th element and $L_2$-norm, respectively.
$\odot$ and $\left<\cdot,\cdot \right>$ denote the element-wise product and the inner product, respectively. 
$\mathrm{j}=\sqrt{-1}$ and $\mathrm{e}$ denote the imaginary unit and the Euler number, respectively.

\section{System Model} \label{sect:system_model}
The system model for SWAN-based ISAC is shown in Fig. \ref{fig:system_model_and_three_protocols}.
The service region of the SWAN-based ISAC system is given by $D_x \times D_y$, where $D_x$ and $D_y$ denote the side length along the $x$- and $y$-axes, respectively.
The base station (BS) is equipped with two segmented waveguides: One for transmission and one for reception.
These waveguides are placed in parallel to the $XOY$ plane at a height of $d$, and extend along the direction of the $x$-axis.
The transmit SWAN (Tx-SWAN) contains $N$ segments, while the receive SWAN (Rx-SWAN) contains $M$ segments.
Each segment contains one PA and is connected to a feed point on the left end, which is illustrated in Fig. \ref{fig:system_model_and_three_protocols}.
There are two types of users considered: $K$ communication users (CUs) and one sensing target (ST).
Geometrically, the $k$-th CU is located at $\mathbf{r}_{\mathrm{c}, k} = [x_{\mathrm{c}, k}, y_{\mathrm{c}, k}, 0]^{\mathrm{T}} \in \mathbb{R}^{3 \times 1}$ with $k \in \mathcal{K} \triangleq \{1,..., K\}$ and the ST is located at $\mathbf{r}_{\mathrm{s}} = [x_{\mathrm{s}}, y_{\mathrm{s}}, 0]^{\mathrm{T}}$, where the PA positions on the $n$-th Tx-SWAN segment and the $m$-th Rx-SWAN segment are denoted as $\mathbf{p}_{\mathrm{t}, n} = [x_{\mathrm{t}, n}, y_{\mathrm{t}}, d]^{\mathrm{T}}$ and $\mathbf{p}_{\mathrm{r}, m} = [x_{\mathrm{r}, m}, y_{\mathrm{r}}, d]^{\mathrm{T}} \in \mathbb{R}^{3 \times 1}$, respectively.
To avoid inter-user interference, $K$ CUs are served via TDMA, while the ST is sensed across all time slots.

\subsection{Channel Model}
For the Tx-SWAN, the PA position of the $n$-th segment is denoted by $\mathbf{p}_{\mathrm{t}, n} = [x_{\mathrm{t}, n}, y_{\mathrm{t}}, d]^{\mathrm{T}}$ with $n \in \mathcal{N} \triangleq \{1,2,...,N\}$.
For the Rx-SWAN, the PA position of the $m$-th segment is denoted by $\mathbf{p}_{\mathrm{r}, m} = [x_{\mathrm{r}, m}, y_{\mathrm{r}}, d]^{\mathrm{T}} \in \mathbb{R}^{3 \times 1}$ with $m \in \mathcal{M} \triangleq \{1,2,...,M\}$ \footnote{As there is a single PA on each segment, we refer ``the $n$-th PA of the Tx-SWAN" simply as ``the $n$-th PA", and ``the $m$-th PA of the Rx-SWAN" simply as ``the $m$-th PA". }.
Then, the feed point at the $n$-th segment of the Tx-SWAN and the $m$-th segment of the Rx-SWAN are given by $\mathbf{p}_{\mathrm{t},n}^{\mathrm{FD}}=[x_{\mathrm{t},n}^{\mathrm{FD}},y_{\mathrm{t}}^{\mathrm{FD}},d]^{\mathrm{T}}$ and $\mathbf{p}_{\mathrm{r},m}^{\mathrm{FD}}=[x_{\mathrm{r},m}^{\mathrm{FD}},y_{\mathrm{r}}^{\mathrm{FD}},d]^{\mathrm{T}}$, respectively.
The intended signals will propagate through two types of channels: in-waveguide and free-space.
Denote $k_{\rm g} \triangleq 2 \pi / \lambda_{\rm g} = 2 \pi / (\lambda_{\rm c}/n_{\mathrm{eff}})$ as the guided wavenumber, where $n_{\mathrm{eff}}$ and $\lambda_{\rm c}$ are the effective refractive index and the wavelength in free space, respectively.
For the in-waveguide channels, the channel coefficients for the $n$-th segment of the Tx-SWAN and the $m$-th segment of the Rx-SWAN are given by 
\begin{align}
    \begin{cases}
	g_{\mathrm{t}, n}\left( x_{\mathrm{t}, n} \right) =10^{-\frac{\kappa}{20}\Delta _{\mathrm{t},n}}\mathrm{e}^{-\mathrm{j}k_{\mathrm{g}}\Delta _{\mathrm{t},n}},\\
	g_{\mathrm{r}, m}\left( x_{\mathrm{r}, m} \right) =10^{-\frac{\kappa}{20}\Delta _{\mathrm{r},m}}\mathrm{e}^{-\mathrm{j}k_{\mathrm{g}}\Delta _{\mathrm{r},m}},\\
    \end{cases} \label{eq:no_inwaveguide_loss}
\end{align}
where $x_{\mathrm{t}, n} \in \mathbb{R}$ and $x_{\mathrm{r}, m} \in \mathbb{R}$ denote the $x$-coordinates of the PAs on the $n$-th Tx-SWAN and the $m$-th Rx-SWAN segments, respectively, $\Delta _{\mathrm{t},n}=|x_{\mathrm{t},n}^{\mathrm{FD}}-x_{\mathrm{t},n}|$ and $\Delta _{\mathrm{r},m}=|x_{\mathrm{r},m}^{\mathrm{FD}}-x_{\mathrm{r},m}^{}|$ denote the propagation distances within the $n$-th Tx-SWAN segment and the $m$-th Rx-SWAN segment, respectively, and $\kappa$ denotes the average attenuation factor along the dielectric waveguide measured in $\mathrm{dB/m}$. 
For the downlink free-space channels, the channel coefficients from the $n$-th PA to the $k$-th CU and the ST are respectively given by
\begin{align}
    \begin{cases}
	h_{\mathrm{c},n, k}\left( x_{\mathrm{t}, n} \right) =\frac{\eta}{r_{\mathrm{c},n,k}}\mathrm{e}^{-\mathrm{j}k_{\mathrm{c}}r_{\mathrm{c},n, k}},\\
	h_{\mathrm{s},n}\left( x_{\mathrm{t}, n} \right) =\frac{\eta}{r_{\mathrm{s},n}}\mathrm{e}^{-\mathrm{j}k_{\mathrm{c}}r_{\mathrm{s},n}},\\
    \end{cases}
\end{align}
where $\eta \triangleq \lambda_{\mathrm{c}} / (4 \pi)$, and $r_{\mathrm{c},n,k}=\left\| \mathbf{p}_{\mathrm{t}, n}-\mathbf{r}_{\mathrm{c}, k} \right\|$ and $r_{\mathrm{s},n}=\left\| \mathbf{p}_{\mathrm{t}, n} -\mathbf{r}_{\mathrm{s}} \right\|$ denote the Euclidean distances between the $n$-th PA of the Tx-SWAN the $k$-th CU and to the ST, respectively.
For the uplink free-space channels, the channel coefficient between the ST and the $m$-th PA is given as follows:
\begin{align}
    f_{\mathrm{s},m}\left( x_{\mathrm{r},m} \right) ={\eta}/{d_{\mathrm{s},m}}\mathrm{e}^{-\mathrm{j}k_{\mathrm{c}}d_{\mathrm{s},m}},
\end{align}
where $d_{\mathrm{s},m}=\left\| \mathbf{p}_{\mathrm{r}, m}-\mathbf{r}_{\mathrm{s}} \right\|$ denotes the Euclidean distance between the $m$-th PA and the ST.

Compactly, the in-waveguide channel vectors for the Tx-SWAN and Rx-SWAN can be respectively defined as 
\begin{align}
	\mathbf{g}_{\mathrm{t}}\left( \mathbf{x}_{\mathrm{t}} \right) &=\left[ g_{\mathrm{t},1}\left( x_{\mathrm{t},1} \right) ,...,g_{\mathrm{t},N}\left( x_{\mathrm{t},N} \right) \right] ^{\mathrm{T}},\\
	\mathbf{g}_{\mathrm{r}}\left( \mathbf{x}_{\mathrm{r}} \right) &=\left[ g_{\mathrm{r},1}\left( x_{\mathrm{r},1} \right) ,...,g_{\mathrm{r},M}\left( x_{\mathrm{r},M} \right) \right] ^{\mathrm{T}},
\end{align}
where $\mathbf{x}_{\mathrm{t}}\triangleq \left[ x_{\mathrm{t},1,...,}x_{\mathrm{t},N} \right]^{\rm T} \in \mathbb{R} ^{N\times 1}$ and $\mathbf{x}_{\mathrm{r}}\triangleq \left[ x_{\mathrm{r},1,...,}x_{\mathrm{r},M} \right]^{\rm T} \in \mathbb{R} ^{M\times 1}$ represent the $x$-coordinate vectors for the Tx-SWAN and the Rx-SWAN, respectively.
For the downlink, the channel vectors for the Tx-SWAN and the Rx-SWAN are respectively defined as
\begin{align}
    	\tilde{\mathbf{h}}_{\mathrm{c},k}\left( \mathbf{x}_{\mathrm{t}} \right) &=\left[ h_{\mathrm{c},1,k}\left( x_{\mathrm{t},1} \right) ,...,h_{\mathrm{c},N,k}\left( x_{\mathrm{t},N} \right) \right] ^{\mathrm{T}},\\
    	\tilde{\mathbf{h}}_{\mathrm{s}}\left( \mathbf{x}_{\mathrm{t}} \right) &=\left[ h_{\mathrm{s},1}\left( x_{\mathrm{t},1} \right) ,...,h_{\mathrm{s},N}\left( x_{\mathrm{t},N} \right) \right] ^{\mathrm{T}}.
\end{align}
For the uplink, the channel vector between the ST and the Rx-SWAN, i.e., free-space channel for echo signals, is defined as
\begin{align}
    \tilde{\mathbf{f}}_{\mathrm{s}}\left( \mathbf{x}_{\mathrm{r}} \right) =\left[ f_{\mathrm{s},1}\left( x_{\mathrm{r},1} \right) ,...,f_{\mathrm{s},M}\left( x_{\mathrm{r},M} \right) \right] ^{\mathrm{T}}.
\end{align}

\subsection{Operating Protocols and Signal Model}
In contrast to conventional PASS, the proposed SWAN architecture may employ three distinct operating protocols, as illustrated in the lower part of Fig. \ref{fig:system_model_and_three_protocols} and summarized as follows:
\begin{itemize}
    \item \textbf{Segment Selection (SS)}: For the SS protocol, only one segment at the Tx-SWAN and one segment at the Rx-SWAN are selected for transmission/reception.
    Both are powered by dedicated RF chains on each side.
    \item \textbf{Segment Aggregation (SA)}: For the SA protocol, all feed points are connected to a shared RF chain on the transmission/reception side, resulting in a natural signal splitting at the Tx and aggregation at the Rx.
    \item \textbf{Segment Multiplexing (SM)}: For the SM protocol, each feed point of the waveguides is connected to a dedicated RF chain on the Tx-SWAN and Rx-SWAN, enabling fully-digital beamforming.
\end{itemize}

In the following paragraphs, we present signal models for all three protocols.
As the CUs are scheduled based on TDMA, we drop the index of the CUs for conciseness.

\subsubsection{SS-Based SWAN ISAC}
In the SS scenario, the received signal at the CU can be expressed as 
\begin{align}
    y_{\mathrm{c}}^{(\mathrm{SS})}=\sqrt{P}\mathbf{h}_{\mathrm{c}}^{\mathrm{T}}\left( \mathbf{x}_{\mathrm{t}} \right) \mathbf{e}_{\mathrm{t}}^{}s_{\mathrm{c}}^{}+n_{\mathrm{c}}^{(\mathrm{SS})}, \label{eq:received_signal_communication_ss}
\end{align}
where $P$ denotes the transmit power, $\mathbf{h}_{\mathrm{c}}^{}\left( \mathbf{x}_{\mathrm{t}} \right) \triangleq \tilde{\mathbf{h}}_{\mathrm{c}}\left( \mathbf{x}_{\mathrm{t}} \right) \odot \mathbf{g}_{\mathrm{t}}^{}\left( \mathbf{x}_{\mathrm{t}} \right)$ denotes the cascaded channel vector for the CU, $s_{\mathrm{c}}^{}$ denotes the communication signal with unit power $\mathbb{E}\{|s_{\mathrm{c}}^{}|^2\}=1$, and $n_{\mathrm{c}} \sim \mathcal{CN}(0, \sigma_{\rm c}^2)$ is the additive Gaussian noise at the CU with $\sigma_{\rm c}^2$ being the noise power.
In \eqref{eq:received_signal_communication_ss}, the selection vector $\mathbf{e}_{\rm t} \in \mathbb{R}^{N \times 1}$ is constrained by $\left\| \mathbf{e}_{\mathrm{t}} \right\|=1$ and $\left[ \mathbf{e}_{\mathrm{t}} \right] _n \in \{0, 1\}$ for $\forall n$, which reflects a one-to-one mapping relationship between the RF chain and the feed points.
For the ST, the communication signal sent by the Tx-SWAN first reaches the ST and then is reflected at the ST.
The echo signal will be received at the Rx-SWAN. 
Hence, the uplink and downlink cascaded channels are given by $\mathbf{f}_{\mathrm{s}}^{}\left( \mathbf{x}_{\mathrm{r}} \right) \triangleq \tilde{\mathbf{f}}_{\mathrm{s}}^{}\left( \mathbf{x}_{\mathrm{r}} \right) \odot \mathbf{g}_{\mathrm{r}}^{}\left( \mathbf{x}_{\mathrm{r}} \right) 
$ and $\mathbf{h}_{\mathrm{s}}^{}\left( \mathbf{x}_{\mathrm{t}} \right) \triangleq \tilde{\mathbf{h}}_{\mathrm{s}}\left( \mathbf{x}_{\mathrm{t}} \right) \odot \mathbf{g}_{\mathrm{t}}^{}\left( \mathbf{x}_{\mathrm{t}} \right) $.
Consequently, assuming perfect self-interference (SI) cancellation, the received echo signal reflected from the ST can be written as
\begin{align}
    y_{\mathrm{s}}^{(\mathrm{SS})}&=\sqrt{\alpha P}\mathbf{e}_{\mathrm{r}}^{\mathrm{H}}\mathbf{f}_{\mathrm{s}}^{}\left( \mathbf{x}_{\mathrm{r}} \right) \mathbf{h}_{\mathrm{s}}^{\mathrm{T}}\left( \mathbf{x}_{\mathrm{t}} \right) \mathbf{e}_{\mathrm{t}}^{}s_{\mathrm{c}}^{} +n_{\mathrm{s}}^{(\mathrm{SS})}, \label{eq:received_signal_sensing}
\end{align}
where $\alpha$ denotes the reflection coefficient at the ST, $n_{\mathrm{s}}^{(\mathrm{SS})} \sim \mathcal{CN}(0, \sigma_{\rm s}^2)$ denotes the additive Gaussian noise at the Rx-SWAN, and $\mathbf{e}_{\mathrm{r}}$ is the one-to-one mapping vector at the Rx-SWAN under the constraint $\left\| \mathbf{e}_{\mathrm{r}} \right\|=1$ and $\left[ \mathbf{e}_{\mathrm{r}} \right] _m \in \{0, 1\}$ for $\forall m$. 
To evaluate communication performance, the data rate is used as a metric.
It is essential to note that inter-user interference does not occur owing to TDMA.
To evaluate this metric, the SNR at the CU can be derived according to \eqref{eq:received_signal_communication_ss}, thus yielding: 
\begin{align}
    \gamma _{\mathrm{c}}^{\left( \mathrm{SS} \right)}= \frac{P | \mathbf{h}_{\mathrm{c}}^{\mathrm{T}}\left( \mathbf{x}_{\mathrm{t}} \right) \mathbf{e}_{\mathrm{t}}^{} |^2}{\sigma _{\mathrm{c}}^{2}}.
\end{align}
Hence, the data rate at the $k$-th CU is given by
\begin{align}
    R_{\mathrm{c}}^{\mathrm{(SS)}}  = \frac{1}{K}\log _2\left( 1+\gamma _{\mathrm{c}}^{\left( \mathrm{SS} \right)} \right). \label{eq:throughput_ss}
\end{align}
The factor $1/K$ is caused by the use of TDMA.
For sensing performance evaluation, we adopt the received SNR of the echo signal in \eqref{eq:received_signal_sensing}.
Here, we assume that the echo signal reflected by the CU can be perfectly canceled out with a successive interference cancellation (SIC) technique.
Thereby, the SNR for the echo signal can be expressed as 
\begin{align}
    \gamma _{\mathrm{s}}^{\left( \mathrm{SS} \right)}=\frac{\alpha P\left| \mathbf{e}_{\mathrm{r}}^{\mathrm{H}}\mathbf{f}_{\mathrm{s}}^{}\left( \mathbf{x}_{\mathrm{r}} \right) \mathbf{h}_{\mathrm{s}}^{\mathrm{T}}\left( \mathbf{x}_{\mathrm{t}} \right) \mathbf{e}_{\mathrm{t}}^{} \right|^2}{\sigma _{\mathrm{s}}^{2}}. \label{eq:sensing_snr_ss}
\end{align}

\subsubsection{SA-Based SWAN ISAC}
Under the SA protocol, all the segments of the Tx-SWAN or Rx-SWAN share the same RF chain. 
Corresponding to this connection topology, the received signal at a CU can be expressed as follows:
\begin{align}
    y_{\mathrm{c}}^{(\mathrm{SA})}= {\sqrt{P /N}} \mathbf{h}_{\mathrm{c}}^{\mathrm{T}}\left( \mathbf{x}_{\mathrm{t}} \right) \boldsymbol{1}_{N}^{}s_{\mathrm{c}}^{}+n_{\mathrm{c}}^{(\mathrm{SA})}, \label{eq:received_signal_communication_sa}
\end{align}
where $n_{\mathrm{c}}^{(\mathrm{SA})} \sim \mathcal{CN}(0, \sigma_{\mathrm{c}}^2)$ denotes the additive Gaussian noise and $\boldsymbol{1}_{N}^{}$ denotes the all-one vector with a dimension of $N$.
In \eqref{eq:received_signal_communication_sa}, as a single RF chain is connected to all segments, we further introduce a scaling factor $1 / \sqrt{N}$ to obey the law of conservation of energy.
For the ST, the received echo signal at the Rx-SWAN can be written as follows:
\begin{align}
    y_{\mathrm{s}}^{(\mathrm{SA})}=\sqrt{\alpha P / N }\boldsymbol{1}_{M}^{\mathrm{T}}\mathbf{f}_{\mathrm{s}}^{}\left( \mathbf{x}_{\mathrm{r}} \right) \mathbf{h}_{\mathrm{s}}^{\mathrm{T}}\left( \mathbf{x}_{\mathrm{t}} \right) \boldsymbol{1}_{N}^{}s_{\mathrm{c}}^{} +n_{\mathrm{s}}^{(\mathrm{SA})}, \label{eq:received_signal_sensing_sa}
\end{align}
where $n_{\mathrm{s}}^{\mathrm{(SA)}} \sim \mathcal{CN}(0, M \sigma_{\rm s}^2)$ denotes the additive Gaussian noise at the Rx-SWAN owing to the noise aggregation of the Rx-SWAN.
For the communication functionality, the SNR at the CU can be computed as
\begin{align}
    \gamma _{\mathrm{c}}^{\left( \mathrm{SA} \right)} =\frac{P|\mathbf{h}_{\mathrm{c}}^{\mathrm{T}}\left( \mathbf{x}_{\mathrm{t}} \right) \boldsymbol{1}_{N}^{} |^2}{N\sigma _{\mathrm{c}}^{2}}.
\end{align}
Correspondingly, the data rate at the CU is given by
\begin{align}
    R_{\mathrm{c}}^{(\mathrm{SA)}} =\frac{1}{K} \log _2\left( 1+\gamma _{\mathrm{c}}^{\left( \mathrm{SA} \right)} \right) .
\end{align}
For the sensing functionality, the received SNR for the echo signal can be derived as
\begin{align}
   \gamma _{\mathrm{s}}^{\left( \mathrm{SA} \right)}=\frac{\alpha P\left| \boldsymbol{1}_{M}^{\mathrm{T}}\mathbf{f}_{\mathrm{s}}^{}\left( \mathbf{x}_{\mathrm{r}} \right) \mathbf{h}_{\mathrm{s}}^{\mathrm{T}}\left( \mathbf{x}_{\mathrm{t}} \right) \boldsymbol{1}_{N}^{} \right|^2}{NM\sigma _{\mathrm{s}}^{2}}.
\end{align}

\subsubsection{SM-Based SWAN ISAC}
In the SM protocol, each segment is connected to a dedicated RF chain for both signal transmission and reception.
Therefore, the received signal at a CU is given by
\begin{align}
    y_{\mathrm{c}}^{(\mathrm{SM)}}=\sqrt{P}\mathbf{h}_{\mathrm{c}}^{\mathrm{H}}\left( \mathbf{x}_{\mathrm{t}} \right) \mathbf{w}_{\mathrm{t}}s_{\mathrm{c}}+n_{\mathrm{c}}^{(\mathrm{SM)}}, \label{eq:received_signal_communication_sm}
\end{align}
where $n_{\mathrm{c}}^{(\rm SM)} \sim \mathcal{CN}(0, \sigma_{\mathrm{c}}^2)$ denotes the additive Gaussian noise at the CU, and $\mathbf{w}_{\mathrm{t}} \in \mathbb{C}^{N \times 1}$ is the beamforming vector.
For the ST, both the transmit beamforming vector and the combining vector need to be considered.
Thereby, the received echo signal reflected from the ST can be formulated as
\begin{align}
    y_{\mathrm{s}}^{(\mathrm{SM)}}=\sqrt{\alpha P}\mathbf{w}_{\mathrm{r}}^{\mathrm{H}}\mathbf{f}_{\mathrm{s}}^{}\left( \mathbf{x}_{\mathrm{r}} \right) \mathbf{h}_{\mathrm{s}}^{\mathrm{H}}\left( \mathbf{x}_{\mathrm{t}} \right) \mathbf{w}_{\mathrm{t}}s_{\mathrm{c},k}+\mathbf{w}_{\mathrm{r}}^{\mathrm{H}}\mathbf{n}_{\mathrm{s}}^{(\mathrm{SM)}}, \label{eq:received_signal_sensing_sm}
\end{align}
where $\|\mathbf{w}_{\mathrm{r}}\|=1$ denotes the combining vector, and $\mathbf{n}_{\mathrm{s}}^{(\mathrm{SM)}}\sim \mathcal{C} \mathcal{N} (0,\sigma _{\mathrm{s}}^{2}\mathbf{I}_M)$ denotes the additive Gaussian noise at the Rx-SWAN.
For communication, the SNR at a CU is given by  
\begin{align}
    \gamma _{\mathrm{c}}^{\left( \mathrm{SM} \right)}=\frac{P|\mathbf{h}_{\mathrm{c}}^{\mathrm{T}}\left( \mathbf{x}_{\mathrm{t}} \right) \mathbf{w}_{\mathrm{t}}|^2}{\sigma _{\mathrm{c}}^{2}}.
\end{align}
Correspondingly, the data rate at this CU is given by
\begin{align}
    R_{\mathrm{c}}^{(\mathrm{SM)}}=\frac{1}{K}\log _2\left( 1+\gamma _{\mathrm{c}}^{\left( \mathrm{SM} \right)} \right) .
\end{align}
For sensing, the received SNR for the echo signal can be expressed as follows:
\begin{align}
    \gamma _{\mathrm{s}}^{\left( \mathrm{SM} \right)}=\frac{\alpha P\left| \mathbf{w}_{\mathrm{r}}^{\mathrm{H}}\mathbf{f}_{\mathrm{s}}^{}\left( \mathbf{x}_{\mathrm{r}} \right) \mathbf{h}_{\mathrm{s}}^{\mathrm{H}}\left( \mathbf{x}_{\mathrm{t}} \right) \mathbf{w}_{\mathrm{t}} \right|^2}{\sigma _{\mathrm{s}}^{2}}.
\end{align}
In what follows, the derived communication and sensing metrics are used to characterize the performance of the proposed ISAC system.

\section{Sensing Performance Limits of SWAN} \label{sect:analystical}
In this section, we first analyze a sensing-centric design, where the primary objective is to greedily maximize the sensing SNR, while the communication functionality is completely disregarded. 
For this design, the sensing superiority of SWAN over conventional PASS is theoretically demonstrated. 

\subsection{Segment Selection}
 As the free-space attenuation for conventional PASS and SWAN is the same, we mainly focus on their differences in terms of the in-waveguide propagation loss.
Assume that the ST is uniformly distributed on the service area, i.e., $x_{\mathrm{s}} \sim \mathcal{U}(0, D_x)$.
The optimal PA positions on the Tx-SWAN and Rx-SWAN are given by $x_{\mathrm{t},n^\star}=x_{\mathrm{r},m^\star}=x_\mathrm{s}$, where $n^\star$ and $m^\star$ denote the selected segments' index on the Tx-SWAN and Rx-SWAN.
As such, the in-waveguide propagation distances within the Tx-SWAN and Rx-SWAN conform to two independent uniform distributions, i.e., $x_{\mathrm{t}, n^\star}^{\mathrm{FD}}-x_{\mathrm{s}} \triangleq \Delta_{x,\mathrm{t}}^{\mathrm{(SWAN)}} \sim \mathcal{U}(0, L)$ and $ x_{\mathrm{r}, m^\star}^{\mathrm{FD}} - x_{\mathrm{s}}  \triangleq \Delta_{x,\mathrm{r}}^{\mathrm{(SWAN)}} \sim \mathcal{U}(0, L)$.
On the contrary, due to the lack of segment structure, the in-waveguide propagation distances for conventional PASS can be specified as follows: $\Delta_{x,\mathrm{t}}^{\mathrm{(PASS)}}  \sim \mathcal{U}(0, D_x)$ and $\Delta_{x,\mathrm{r}}^{\mathrm{(PASS)}} \sim \mathcal{U}(0, D_x)$ for the transmitter and receiver sides, respectively.
Therefore, the in-waveguide propagation gains of SWAN-based sensing and conventional PASS-based sensing can be expressed as follows:
\begin{align}
    &\gamma _{\mathrm{SWAN}}^{\left( \mathrm{SS} \right)}=\int_0^L{\int_0^L{ \frac{1}{L^2}10^{-\frac{\kappa}{10}x_{\mathrm{t}}}10^{-\frac{\kappa}{10}x_{\mathrm{r}}}}\mathrm{d}x_{\mathrm{t}}}\mathrm{d}x_{\mathrm{r}},
    \label{eq:gain_ss_swan}\\
    &\gamma _{\mathrm{PASS}}^{\left( \mathrm{SS} \right)}=\int_0^{D_x}{\int_0^{D_x}{\frac{1}{D_x^2}10^{-\frac{\kappa}{10}x_{\mathrm{t}}}10^{-\frac{\kappa}{10}x_{\mathrm{r}}}}\mathrm{d}x_{\mathrm{t}}}\mathrm{d} x_{\mathrm{r}}. \label{eq:gain_ss_pass}
\end{align}
Further, expressions \eqref{eq:gain_ss_swan} and \eqref{eq:gain_ss_pass} can be calculated as
\begin{align}
    \gamma _{\mathrm{SWAN}}^{\left( \mathrm{SS} \right)}&=\frac{\left( 1-{\mathrm{e}}^{-2\beta D_x/M} \right) \left( 1-{\mathrm{e}}^{-2\beta D_x/N} \right)}{4\beta ^2D_{x}^{2}/\left( MN \right)}, \\
    \gamma _{\mathrm{PASS}}^{\left( \mathrm{SS} \right)}&=\frac{\left( 1-{\mathrm{e}}^{-2\beta D_x} \right) ^2}{4\beta ^2D_{x}^{2}},
\end{align}
where $\beta \triangleq \kappa \ln 10/20$.
Moreover, it is easy to verify that $\lim_{M, N \rightarrow +\infty} \gamma _{\mathrm{SWAN}}^{\left( \mathrm{SS} \right)} =1 $, which implies that there will be no in-waveguide loss as the segment's length reaches zero.
The ratio of the in-waveguide losses representing the gain of SWAN over PASS is given by
\begin{align}
    \eta^{\mathrm{(SS)}} =\frac{\gamma _{\mathrm{SWAN}}^{\left( \mathrm{SS} \right)}}{\gamma _{\mathrm{PASS}}^{\left( \mathrm{SS} \right)}}=\frac{MN\left( 1-{\mathrm{e}}^{-2\beta D_x/M} \right) \left( 1-{\mathrm{e}}^{-2\beta D_x/N} \right)}{\left( 1-{\mathrm{e}}^{-2\beta D_x} \right) ^2}. \notag 
\end{align}
From an asymptotic perspective, an upper bound of $\eta^{\mathrm{(SS)}}$ is given as follows:
\begin{align}
    \underset{M,N\rightarrow +\infty}{\lim}\eta^{\mathrm{(SS)}} =\frac{4\beta ^2D_{x}^{2}}{\left( 1-{\mathrm{e}}^{-2\beta D_x} \right) ^2}.
\end{align}
\begin{remark}\emph{\emph{(Superiority of SWAN Sensing Under SS)} Owning to the segmented structure of SWAN, the in-waveguide loss is less severe, thereby contributing to SWAN's gain over conventional PASS. 
As the number of segments increases, this gain increases correspondingly.
When the number of segments goes to infinity, the ultimate gain is determined solely by the side length $D_x$ (i.e., the width of the service area).}
\end{remark}

\subsection{Segment Aggregation}
Compared with the SS protocol, the SA protocol requires optimizing multiple PA positions for each waveguide. 
Accordingly, a ``coarse-then-refine” approach is adopted: first, coarse PA positions are determined to minimize the path loss; then, these positions are refined to align phase shifts, achieving constructive superposition.
It is noted that the pathloss changes caused by the refine step are negligible, as this step only induces wavelength-scale pathloss variations \footnote{To maintain a balanced presentation across sections, the detailed steps of the ``coarse-then-refine” procedure are deferred to the next section. Here, we only present the sensing SNR obtained with this approach. For the implementation details, please refer to Section \ref{sect:sa_based_isac_single_cu}.}.
Define $\Delta _{\mathrm{t},\mathrm{s}}^{2}\triangleq {\left( y_{\mathrm{t}}-y_{\mathrm{s}} \right) ^2+d^2}
$ and $\Delta _{\mathrm{r},\mathrm{s}}^{2}\triangleq {\left( y_{\mathrm{r}}-y_{\mathrm{s}} \right) ^2+d^2}$.
The sensing SNR for the SA protocol is given by
\begin{align}
    &\gamma _{\mathrm{s}}^{\left( \mathrm{SA} \right)}=\frac{\alpha P\eta ^4}{NM\sigma _{\mathrm{s}}^{2}}\left| \sum_{n=1}^N{\frac{\mathrm{e}^{-\mathrm{j}k_{\mathrm{c}}\sqrt{\left( \tilde{x}_{\mathrm{t},n}-x_{\mathrm{s}} \right) ^2+\Delta _{\mathrm{t},\mathrm{s}}^{2}}}\mathrm{e}^{-\mathrm{j}k_{\mathrm{g}}\Delta _{\,\,\mathrm{t},n}}}{\sqrt{\left( \tilde{x}_{\mathrm{t},n}-x_{\mathrm{s}} \right) ^2+\Delta _{\mathrm{t},\mathrm{s}}^{2}}}} \right|^2  \notag \\
    &\times \left| \sum_{m=1}^M{\frac{\mathrm{e}^{-\mathrm{j}k_{\mathrm{c}}\sqrt{\left( \tilde{x}_{\mathrm{r},m}-x_{\mathrm{s}} \right) ^2+\Delta _{\mathrm{r},\mathrm{s}}^{2}}}\mathrm{e}^{-\mathrm{j}k_{\mathrm{g}}\Delta _{\mathrm{r},n}}}{\sqrt{\left( \tilde{x}_{\mathrm{r},m}-x_{\mathrm{s}} \right) ^2+\Delta _{\mathrm{r},\mathrm{s}}^{2}}}} \right|^2 \notag
\\
&\overset{\mathrm{(a)}}{=}\frac{\alpha P\eta ^4}{NM\sigma _{\mathrm{s}}^{2}}\left| \sum_{n=1}^N{\frac{1}{\sqrt{\left( \tilde{x}_{\mathrm{t},n}-x_{\mathrm{s}} \right) ^2+\Delta _{\mathrm{t},\mathrm{s}}^{2}}}} \right|^2 \notag \\
&\qquad \qquad \qquad \times \left| \sum_{m=1}^M{\frac{1}{\sqrt{\left( \tilde{x}_{\mathrm{r},m}-x_{\mathrm{s}} \right) ^2+\Delta _{\mathrm{r},\mathrm{s}}^{2}}}} \right|^2
\notag \\
&\overset{\left( \mathrm{b} \right)}{\approx}\frac{\alpha P\eta ^4}{NM\sigma _{\mathrm{s}}^{2}}\underset{\triangleq A^{\left( \mathrm{SA} \right)}(\mathbf{x}_{\mathrm{t}}^{})} {\underbrace{\left| \sum_{n=1}^N{\frac{1}{\sqrt{\left( x_{\mathrm{t},n}-x_{\mathrm{s}} \right) ^2+\Delta _{\mathrm{t},\mathrm{s}}^{2}}}} \right|^2}} \notag \\
&\qquad \qquad \qquad \times \underset{\triangleq B^{\left( \mathrm{SA} \right)}(\mathbf{x}_{\mathrm{r}}^{})}{\underbrace{\left| \sum_{m=1}^M{\frac{1}{\sqrt{\left( x_{\mathrm{r},m}-x_{\mathrm{s}} \right) ^2+\Delta _{\mathrm{r},\mathrm{s}}^{2}}}} \right|^2}}
, \label{eq:gamma_sa_max_approx}
\end{align}
where we introduced $\tilde{x}_{\mathrm{t},n}=x_{\mathrm{t},n}+\Delta _{x_{\mathrm{t},n}}$ and $\tilde{x}_{\mathrm{r},n}=x_{\mathrm{r},n}+\Delta _{x_{\mathrm{r},n}}$, with $\Delta _{x_{\mathrm{t},n}}$ and $\Delta _{x_{\mathrm{r},n}}$ being the lengths of the refinement step for the Tx-SWAN and Rx-SWAN, respectively.
By judiciously designing $\Delta _{x_{\mathrm{t},n}}$ and $\Delta _{x_{\mathrm{r},n}}$, step (a) is achieved by aligning the phases of all the PAs to ensure constructive superpositions for downlink probing transmission and uplink echo reception, respectively.
Furthermore, the lengths of the refinement steps are on the order of the wavelength, and due to the high carrier frequency used for SWANs, the resulting impact on large-scale fading is negligible \cite{ouyang2025uplink}.
Thus, the approximation in step $\rm (b)$ is achieved by neglecting the pathloss changes induced by the refinement step.
The maximal sensing SNR can be expressed as follows:
\begin{align}
    \gamma _{\mathrm{SWAN}}^{\left( \mathrm{SA} \right)}=\max\nolimits_{\mathbf{x}_{\mathrm{t}}^{},\mathbf{x}_{\mathrm{r}}^{}}\gamma _{\mathrm{s}}^{\left( \mathrm{SA} \right)}.
\end{align}
Under the assumption that the ST is located at the center of the service area.
We can obtain the optimal PA position vectors $\mathbf{x}_{\mathrm{t}}^{\star}$ and $\mathbf{x}_{\mathrm{r}}^{\star}$ as follows: Set the $x$-coordinates of the PAs on the middle segments equal to the $x$-coordinate of the service area center; place the left-side PAs (relative to the waveguide centers) at the segment endpoints, and the right-side PAs at the segment feed points, so as to minimize the pathloss. 
As such, owing to the symmetric placement of the PAs, terms $A^{\left( \mathrm{SA} \right)}(\mathbf{x}_{\mathrm{t}}^{\star})$ and $B^{\left( \mathrm{SA} \right)}(\mathbf{x}_{\mathrm{r}}^{\star})$ can be simplified as follows:
\begin{align}
    &A^{\left( \mathrm{SA} \right)}(\mathbf{x}_{\mathrm{t}}^{\star})=\left| \frac{1}{\Delta _{\mathrm{t},\mathrm{s}}^{}}+\sum_{\tilde{n}=1}^{\frac{N-1}{2}}{\frac{2}{\sqrt{L^2\left( \tilde{n}-1/2 \right) ^2+\Delta _{\mathrm{t},\mathrm{s}}^{2}}}} \right|^2, \notag \\
    &B^{\left( \mathrm{SA} \right)}(\mathbf{x}_{\mathrm{r}}^{\star})=\left| \frac{1}{\Delta _{\mathrm{r},\mathrm{s}}^{}}+\sum_{\tilde{m}=1}^{\frac{M-1}{2}}{\frac{2}{\sqrt{L^2\left( \tilde{m}-1/2 \right) ^2+\Delta _{\mathrm{r},\mathrm{s}}^{2}}}} \right|^2. \notag
\end{align}
Further, by adopting \cite[Lemma 2]{ouyang2025uplink}, for large $M$, the terms $A^{\left( \mathrm{SA} \right)}(\mathbf{x}_{\mathrm{t}}^{\star})$ and $B^{\left( \mathrm{SA} \right)}(\mathbf{x}_{\mathrm{r}}^{\star})$ can be simplified into the following closed-form expressions:
\begin{align}
    A^{\left( \mathrm{SA} \right)}(\mathbf{x}_{\mathrm{t}}^{\star})&\simeq \left( \frac{1}{\Delta _{\mathrm{t},\mathrm{s}}^{}}+\frac{2N}{D_x}\sinh ^{-1}\left( \frac{D_x}{2\Delta _{\mathrm{t},\mathrm{s}}^{}} \right) \right) ^2,
    \\
    B^{\left( \mathrm{SA} \right)}(\mathbf{x}_{\mathrm{r}}^{\star})&\simeq \left( \frac{1}{\Delta _{\mathrm{r},\mathrm{s}}^{}}+\frac{2M}{D_x}\sinh ^{-1}\left( \frac{D_x}{2\Delta _{\mathrm{r},\mathrm{s}}^{}} \right) \right) ^2,
\end{align}
where the approximation $\left( M-1 \right) L=\left( N-1 \right) L\simeq ML=NL=D_x$ is used.
In comparison, for conventional PASS, the echo signal SNR can be expressed as $\gamma _{\mathrm{PASS}}^{\left( \mathrm{SA} \right)}\simeq \frac{\alpha P\eta ^4}{\sigma _{\mathrm{s}}^{2}\Delta _{\mathrm{t},\mathrm{s}}^{2}\Delta _{\mathrm{r},\mathrm{s}}^{2}}$.
Therefore, the gain achieved by adopting SWAN over conventional PASS can be derived as
\begin{align}
    \eta ^{(\mathrm{SA)}}\triangleq \frac{\gamma _{\mathrm{SWAN}}^{\left( \mathrm{SA} \right)}}{\gamma _{\mathrm{PASS}}^{\left( \mathrm{SA} \right)}}&=\frac{\Delta _{\mathrm{t},\mathrm{s}}^{2}\Delta _{\mathrm{r},\mathrm{s}}^{2}}{NM}\left( \frac{1}{\Delta _{\mathrm{t},\mathrm{s}}^{}}+\frac{2N}{D_x}\sinh ^{-1}\left( \frac{D_x}{2\Delta _{\mathrm{t},\mathrm{s}}^{}} \right) \right) ^2 \notag \\
    &\times \left( \frac{1}{\Delta _{\mathrm{r},\mathrm{s}}^{}}+\frac{2M}{D_x}\sinh ^{-1}\left( \frac{D_x}{2\Delta _{\mathrm{r},\mathrm{s}}^{}} \right) \right) ^2. 
\end{align}
Asymptotically, this gain can be expressed as follows:
\begin{align}
    \lim_{M,N\rightarrow \infty} \eta ^{(\mathrm{SA)}}&=\frac{16\Delta _{\mathrm{t},\mathrm{s}}^{2}\Delta _{\mathrm{r},\mathrm{s}}^{2}NM}{D_{x}^{4}}\left( \sinh ^{-1}\left( \frac{D_x}{2\Delta _{\mathrm{r, s}}} \right) \right) ^2 \notag \\
    &\quad \quad \quad \times \left( \sinh ^{-1}\left( \frac{D_x}{2\Delta _{\mathrm{t, s}}} \right) \right) ^2.
\end{align}
\begin{remark}\emph{\emph{(Superiority of SWAN Sensing Under SA)} Under the SA scenario, the sensing SNR initially decreases with $N$ and $M$.
After $N^{\star}=\frac{D_x}{2\Delta _{\mathrm{t},\mathrm{s}}^{}\sinh ^{-1}\left( D_x/\left( 2\Delta _{\mathrm{t},\mathrm{s}}^{} \right) \right)}$ and $M^{\star}=\frac{D_x}{2\Delta _{\mathrm{r},\mathrm{s}}^{}\sinh ^{-1}\left( D_x/\left( 2\Delta _{\mathrm{r},\mathrm{s}}^{} \right) \right)}$ are reached, the sensing SNR increases monotonically.
When the number of segments goes to infinity, the ultimate gain scales as $\mathcal{O}(MN)$ with a linear coefficient determined by $D_x$.}
\end{remark}

\subsection{Segment Multiplexing}
In this sub-section, we consider the sensing-centric design for the SM protocol.
Compared with SA, SM allows beamforming on top of PA repositioning.
To maximize the sensing SNR, maximum ratio transmission (MRT) and maximum ratio combination (MRC) can be utilized at the Tx-SWAN and Rx-SWAN, respectively.
Owing to the dedicated RF chains for the segments, the phase-shift alignment, previously achieved via the ``coarse-then-fine” method in the SA protocol, can now be automatically realized under the SM protocol.
In this case, the sensing SNR of the echo signal is given by
\begin{align}
    &\gamma _{\mathrm{s}}^{\left( \mathrm{SM} \right)}=\frac{\alpha P\eta ^4}{\sigma _{\mathrm{s}}^{2}}\sum_{n=1}^N{\frac{1}{\left( {x}_{\mathrm{t},n}-x_{\mathrm{s}} \right) ^2+\Delta _{\mathrm{t},\mathrm{s}}^{2}}}\notag \\
    &\times {\sum_{m=1}^M{\frac{1}{\left( {x}_{\mathrm{r},m}-x_{\mathrm{s}} \right) ^2+\Delta _{\mathrm{r},\mathrm{s}}^{2}}}}=\frac{\alpha P\eta ^4}{\sigma _{\mathrm{s}}^{2}}A^{\left( \mathrm{SM} \right)}(\mathbf{x}_{\rm t})B^{\left( \mathrm{SM} \right)}(\mathbf{x}_{\rm r}).
\end{align}
Hence, the maximum sensing SNR is given by
\begin{align}
     \gamma _{\mathrm{SWAN}}^{\left( \mathrm{SM} \right)}=\max\nolimits_{\mathbf{x}_{\mathrm{t}}^{},\mathbf{x}_{\mathrm{r}}^{}}\gamma _{\mathrm{s}}^{\left( \mathrm{SM} \right)}.
\end{align}
Similar to the analysis under the SA protocol, we assume that the ST is located at the center of the service area, which will give rise to the following simplified forms:
\begin{align}
    &A^{\left( \mathrm{SM} \right)}(\mathbf{x}_{\mathrm{t}}^{\star})= \frac{1}{\Delta _{\mathrm{t},\mathrm{s}}^{2}}+\sum_{\tilde{n}=1}^{\frac{N-1}{2}}{\frac{2}{{L^2\left( \tilde{n}-1/2 \right) ^2+\Delta _{\mathrm{t},\mathrm{s}}^{2}}}}, \\
    &B^{\left( \mathrm{SM} \right)}(\mathbf{x}_{\mathrm{r}}^{\star})= \frac{1}{\Delta _{\mathrm{r},\mathrm{s}}^{2}}+\sum_{\tilde{m}=1}^{\frac{M-1}{2}}{\frac{2}{{L^2\left( \tilde{m}-1/2 \right) ^2+\Delta _{\mathrm{r},\mathrm{s}}^{2}}}}.
\end{align}
In the above equations, the optimal PA positions, i.e., $\mathbf{x}_{\mathrm{t}}^\star$ and $\mathbf{x}_{\mathrm{r}}^\star$, follow the same placement strategy as in the SA protocol to minimize pathloss, while phase alignment is instead achieved through MRT and MRC enabled by the dedicated RF chains.
Thus, according to \cite{ouyang2025uplink}, for a large number of segments, $ \gamma _{\mathrm{SWAN}}^{\left( \mathrm{SM} \right)}$ can be approximated as
\begin{align}
    &\gamma _{\mathrm{SWAN}}^{\left( \mathrm{SM} \right)}\simeq \frac{\alpha P\eta ^4}{\sigma _{\mathrm{s}}^{2}}\left( \frac{1}{\Delta _{\mathrm{r},\mathrm{s}}^{2}}+\frac{2}{L\Delta _{\mathrm{r},\mathrm{s}}^{}}\tan ^{-1}\left( \frac{\left( N-1 \right) L}{2\Delta _{\mathrm{r},\mathrm{s}}^{}} \right) \right) \notag \\
    &\qquad \qquad \times \left( \frac{1}{\Delta _{\mathrm{t},\mathrm{s}}^{2}}+\frac{2}{L\Delta _{\mathrm{t},\mathrm{s}}^{}}\tan ^{-1}\left( \frac{\left( M-1 \right) L}{2\Delta _{\mathrm{t},\mathrm{s}}^{}} \right) \right).
\end{align}
Thereby, the SWAN's gain over conventional PASS under the SM protocol can be derived as follows:
\begin{align}
    \eta ^{(\mathrm{SM)}}&=\frac{\gamma _{\mathrm{SWAN}}^{\left( \mathrm{SM} \right)}}{\gamma _{\mathrm{PASS}}^{\left( \mathrm{SM} \right)}}=\left( 1+\frac{2\Delta _{\mathrm{r},\mathrm{s}}^{}}{L}\tan ^{-1}\left( \frac{\left( N-1 \right) L}{2\Delta _{\mathrm{r},\mathrm{s}}^{}} \right) \right) \notag \\
    &\times \left( 1+\frac{2\Delta _{\mathrm{t},\mathrm{s}}^{}}{L}\tan ^{-1}\left( \frac{\left( M-1 \right) L}{2\Delta _{\mathrm{t},\mathrm{s}}^{}} \right) \right) .
\end{align}
Moreover, as $M$ and $N$ go to infinity, an asymptotic expression of this ratio can be obtained as 
\begin{align}
    \eta ^{(\mathrm{SM)}}&\simeq \left( 1+\frac{2\Delta _{\mathrm{r},\mathrm{s}}^{}}{L}\tan ^{-1}\left( \frac{D_x}{2\Delta _{\mathrm{r},\mathrm{s}}^{}} \right) \right) \notag \\
    &\qquad \quad \times \left( 1+\frac{2\Delta _{\mathrm{t},\mathrm{s}}^{}}{L}\tan ^{-1}\left( \frac{D_x}{2\Delta _{\mathrm{t},\mathrm{s}}^{}} \right) \right).
\end{align}
\begin{remark}\emph{\emph{(Superiority of SWAN Sensing Under SM)} Under the SM protocol, the sensing SNR increases with $N$ and $M$.
When the number of segments approaches infinity, the ultimate gain is bounded by the service area’s side length $D_x$ in the direction of the waveguide.}
\end{remark}

\section{Pareto Front in SWAN-Assisted ISAC Systems} \label{sect:two-user}
In this section, we analyze the special case with one CU and one ST to gain insights into the sensing and communication tradeoff.
Moreover, we compare the three protocols to highlight their respective benefits and costs.

\subsection{SS-Based ISAC}
Since there is only one CU, the subscript $k$ is omitted.
Under the SS protocol, the antenna positions must be determined on the selected segments of the Tx-SWAN and Rx-SWAN.
Thus, the optimization problem is formulated as follows:
\begin{problem}\label{pb:ss-two-user-1} 
  \begin{alignat}{2}
    \underset{x_{\mathrm{t},n},x_{\mathrm{r},m},n,m}{\max} &\quad \log _2( 1+\gamma _{\mathrm{c}}^{\left( \mathrm{SS} \right)}) 
 \label{obj:ss-two-user-1}
    \\ \mathrm{s.t.} &\quad  \gamma _{\mathrm{s}}^{\left( \mathrm{SS} \right)}\ge \Gamma _{\mathrm{sen}}  \label{con:ss-two-user-1a}
    \\  &\quad L(n-1) \leq x_{\mathrm{t},n} \leq Ln \label{con:ss-two-user-1c}
    \\  &\quad L(m-1) \leq x_{\mathrm{r},m} \leq Lm. \label{con:ss-two-user-1d}
   \end{alignat}
\end{problem} 
In problem \eqref{pb:ss-two-user-1}, the objective is to determine the segment indices $n$ and $m$ and the corresponding PA positions $x_{\mathrm{t},n}$ and $x_{\mathrm{r},m}$ within the selected segments, so that the communication rate is maximized while the sensing constraint is met.
To address this problem, we first present the following lemma to determine the optimal segment selection on the Rx-SWAN.
\begin{lemma} \label{lemma:1}
    \normalfont The optimal segment index on the Rx-SWAN under the SS protocol is given by
    \begin{align}
    m^{\star}=\lceil (x_{\mathrm{s}}-x_{\mathrm{r},1}^{\mathrm{FD}})/L \rceil ,\quad x_{\mathrm{r},m^{\star}}=x_s, 
    \end{align}
where $x_{\mathrm{r},1}^{\mathrm{FD}}$ denotes the $x$-coordinate of the feed point on the first segment of the Rx-SWAN.
\end{lemma}
\begin{IEEEproof}
This lemma is established employing proof by contradiction.
In particular, selecting a segment closer to the ST can further increasing the sensing SNR, leading to $ \gamma _{\mathrm{s}, +}^{\left( \mathrm{SS} \right)}\ge \gamma _{\mathrm{s}}^{\left( \mathrm{SS} \right)} \ge \Gamma _{\mathrm{sen}}$, where $\gamma _{\mathrm{s}, +}^{\left( \mathrm{SS} \right)}$ denotes the enhanced echo SNR obtained by choosing a closer segment to the ST and $\gamma _{\mathrm{s}}^{\left( \mathrm{SS} \right)}$ denotes the original one.
Consequently, a better feasible solution can be constructed by choosing a PA closer to the CU, which will be accompanied by a lower sensing SNR at the ST, until $\gamma _{\mathrm{s}, +}^{\left( \mathrm{SS} \right)} = \gamma _{\mathrm{s}}^{\left( \mathrm{SS} \right)}$.
\end{IEEEproof}
Given that $x_{\mathrm{r}, m^\star}$ and $m^\star$ are fixed, we now identify the optimal solution to $x_{\mathrm{t}, n^\star}$ and $n^\star$.
To simplify the optimization process, we omit the segment index $n$, and denote the $x$-coordinate of the PA as $x_{\mathrm{t}}$, indicating it can be repositioned across the whole waveguide, i.e., $[0, D_x]$. 
Once the optimal position $x_{\mathrm{t}}^\star$ is found, the segment covering this position will be selected correspondingly to obtain $m^\star$.
Therefore, the resulting optimization problem is stated as follows:
\begin{problem}\label{pb:ss-two-user-2} 
  \begin{alignat}{2}
    \underset{ 0\le x_{\mathrm{t}} \le D_x}{\max} & \quad \eta ^2P/\sigma _{\mathrm{c}}^{2}\left( \left( x_{\mathrm{t}}-x_{\mathrm{c}} \right) ^2+\Delta _{\mathrm{t},\mathrm{c}}^{2} \right)
 \label{obj:ss-two-user-2}
    \\ \mathrm{s.t.} & \quad \frac{\alpha \eta ^4P_{\max}}{\sigma _{\mathrm{s}}^{2}\left( \left( x_{\mathrm{t}}-x_{\mathrm{s}} \right) ^2+\Delta _{\mathrm{t},\mathrm{s}}^{2} \right) \Delta _{\mathrm{r},\mathrm{s}}^{2}}\ge \Gamma _{\mathrm{sen}}. \label{con:ss-two-user-2a}
   \end{alignat}
\end{problem} 
In problem \eqref{pb:ss-two-user-2}, we define $\Delta _{\mathrm{t},\mathrm{s}}^{2}\triangleq \left( y_{\mathrm{t}}-y_{\mathrm{s}} \right) ^2+d^2$, $\Delta _{\mathrm{r},\mathrm{s}}^{2}\triangleq \left( y_{\mathrm{r}}-y_{\mathrm{s}} \right) ^2+d^2$, $\Delta _{\mathrm{t},\mathrm{c}}^{2}\triangleq \left( y_{\mathrm{t}}-y_{\mathrm{c}} \right) ^2+d^2$, and $\Delta _{\mathrm{r},\mathrm{c}}^{2}\triangleq \left( y_{\mathrm{r}}-y_{\mathrm{c}} \right) ^2+d^2$.
According to \cite{ouyang2025uplink}, the in-waveguide loss can be neglected when only a few segments are used at the transceivers. 
Hence, we ignore this loss for analytical tractability and later validate the tightness of this assumption through simulations.
Regarding \eqref{pb:ss-two-user-2}, as $x_{\mathrm{t}} \rightarrow x_{\mathrm{c}}$, the objective function increases, whereas the sensing SNR reaches its maximum at $x_{\mathrm{t}}=x_{\mathrm{s}}$ and decreases if $x_{\mathrm{t}}$ deviates from this point.
Therefore, the optimal PA position on the Tx-SWAN is the one closest to $x_{\mathrm{c}}$ while satisfying the sensing threshold. 
Accordingly, we first derive the feasible region of $x_{\mathrm{t}}$ that meets constraint~\eqref{con:ss-two-user-2a}, yielding:
\begin{align}
    \overset{\triangleq x_{\mathrm{t}}^{-}}{\overbrace{x_{\mathrm{s}}-\sqrt{\frac{\eta ^4\alpha P_{\max}}{\sigma _{\mathrm{s}}^{2}\Gamma _{\mathrm{sen}}\Delta _{\mathrm{r},\mathrm{s}}^{2}}-\Delta _{\mathrm{t},\mathrm{s}}^{2}}}}\le x_{\mathrm{t}}\le \overset{\triangleq x_{\mathrm{t}}^{+}}{\overbrace{x_{\mathrm{s}}+\sqrt{\frac{\eta ^4\alpha P_{\max}}{\sigma _{\mathrm{s}}^{2}\Gamma _{\mathrm{sen}}\Delta _{\mathrm{r},\mathrm{s}}^{2}}-\Delta _{\mathrm{t},\mathrm{s}}^{2}}}}. \notag
\end{align}
Considering the above, the feasible set of $x_{\mathrm{t}}$ can be written as
\begin{align}
    x_{\mathrm{t}} \in \mathcal{F} \triangleq [\max\{0, x_{\mathrm{t}}^-\}, \min\{x_{\mathrm{t}}^+, D_x \}].
\end{align}
Therefore, the optimal PA placement can be derived as 
\begin{align}
    x_{\mathrm{t}}^{\star}=\begin{cases}
	x_{\mathrm{c}},&		\mathrm{if}~ x_{\mathrm{c}}\in \mathcal{F} ,\\
	\mathrm{arg}\min_{x\in \mathcal{S}}\left| x_{\mathrm{c}}-x \right|\,\,,&		\mathrm{else},\\
\end{cases}
\end{align}
where $\mathcal{S} \triangleq \{\max\{0, x_{\mathrm{t}}^-\}, \min\{x_{\mathrm{t}}^+, D_x \}\}$ denotes the search set, containing the end points of the feasible set.
According to $x_{\mathrm{t}}^{\star}$, the index of the selected segment at Tx-SWAN is 
\begin{align}
    n^{\star}=\lceil (x_{\mathrm{t}}^{\star}-x_{\mathrm{t},1}^{\mathrm{FD}})/L \rceil.
\end{align}
Once the optimal positions for the PAs on the Tx-SWAN and Rx-SWAN have been obtained, the sensing-throughput tradeoff can be revealed by setting different sensing thresholds $\Gamma_{\mathrm{sen}}$.
Since the optimal PA positions are obtained in closed form, the computational complexity for solving the SS problem is $\mathcal{O}(1)$.

\subsection{SA-Based ISAC} \label{sect:sa_based_isac_single_cu}
Compared to the SS protocol, now two $x$-coordinate vectors have to be optimized: One for the Tx-SWAN and one for the Rx-SWAN.
Hence, the optimization problem can be formulated as follows:
\begin{problem}\label{pb:sa-two-user-1} 
  \begin{alignat}{2}
    \underset{\mathbf{x}_{\mathrm{t}}, \mathbf{x}_{\mathrm{r}}}{\max} &\quad  \log _2\left( 1+\gamma _{\mathrm{c}}^{\left( \mathrm{SA} \right)} \right) \label{obj:sa-two-user-1}
    \\ \mathrm{s.t.} &\quad  \gamma _{\mathrm{s}}^{\left( \mathrm{SA} \right)}\ge \Gamma _{\mathrm{sen}}  \label{con:sa-two-user-1a}
    \\  &\quad \mathbf{x}_{\mathrm{t}} \in \mathcal{F}_{\mathrm{t}}~\text{and}~\mathbf{x}_{\mathrm{r}} \in \mathcal{F}_{\mathrm{r}}, \label{con:sa-two-user-1b}
   \end{alignat}
\end{problem} 
where feasible sets $\mathcal{F}_{\mathrm{t}}$ and $\mathcal{F}_{\mathrm{r}}$ are specified as
\begin{align}
    \mathcal{F} _{\mathrm{t}}&\triangleq \left\{ \mathbf{x}_{\mathrm{t}}\left| \begin{array}{c}
	(n-1)L\le [\mathbf{x}_{\mathrm{t}}]_n\le nL,\forall n\in \mathcal{N}\\
	\left| [\mathbf{x}_{\mathrm{t}}]_n-[\mathbf{x}_{\mathrm{t}}]_{n^{\prime}} \right|\ge \Delta _{\min},\forall n\ne n^{\prime}\\
\end{array} \right. \right\}, \\
\mathcal{F} _{\mathrm{r}}&\triangleq \left\{ \mathbf{x}_{\mathrm{r}}\left| \begin{array}{c}
	(m-1)L\le [\mathbf{x}_{\mathrm{r}}]_m\le mL,\forall m\in \mathcal{M}\\
	\left| [\mathbf{x}_{\mathrm{r}}]_m-[\mathbf{x}_{\mathrm{r}}]_{m^{\prime}} \right|\ge \Delta _{\min},\forall m\ne m^{\prime}\\
\end{array} \right. \right\}.
\end{align}
In the above feasible sets, $\Delta_{\min}$ represents the minimum spacing between two adjacent PAs to circumvent mutual coupling effects.
In contrast to \eqref{pb:ss-two-user-1} in the previous section, all segments are activated simultaneously.
To resolve this problem, we first rewrite the optimization objective \eqref{obj:sa-two-user-1} as follows:
\begin{align}
    \eqref{obj:sa-two-user-1} = \log _2\left( 1+\left| \sum_{n=1}^N{({\eta P} /{(\sigma_{\mathrm{c}}^2r_{\mathrm{c},n})})\mathrm{e}^{-\mathrm{j}\left( k_{\mathrm{c}}r_{\mathrm{c},n}+k_{\mathrm{g}}\Delta _{\mathrm{t},n} \right)}} \right|^2 \right). \notag 
\end{align}
Therefore, for simplicity, instead of optimizing \eqref{obj:sa-two-user-1} directly, we can optimize the SNR term at CU, i.e., $\gamma _{\mathrm{c}}^{\left( \mathrm{SA} \right)}\left( \mathbf{x}_{\mathrm{t}} \right) \triangleq | \sum_{n=1}^N{({\eta P} /{(\sigma_{\mathrm{c}}^2r_{\mathrm{c},n})})\mathrm{e}^{-\mathrm{j}\left( k_{\mathrm{c}}r_{\mathrm{c},n}+k_{\mathrm{g}}\Delta _{\mathrm{t},n} \right)}}|^2$, owning to their equivalence.
For the constraint, the sensing threshold constraint \eqref{con:sa-two-user-1a} can also be written as
\begin{align}
    &\gamma _{\mathrm{s}}^{\left( \mathrm{SA} \right)}\left( \mathbf{x}_{\mathrm{t}},\mathbf{x}_{\mathrm{r}} \right)=\frac{\alpha P  \eta^4}{NM\sigma _{\mathrm{s}}^{2}} \times 
    \notag \\
    &\left| \sum_{n=1}^N{\frac{\eta}{r_{\mathrm{s},n}}\mathrm{e}^{-\mathrm{j}\left( k_{\mathrm{c}}r_{\mathrm{s},n}+k_{\mathrm{g}}\Delta _{\mathrm{t},n} \right)}} \right|^2\left| \sum_{m=1}^M{\frac{\eta}{d_{\mathrm{s},m}}\mathrm{e}^{-\mathrm{j}\left( k_{\mathrm{c}}d_{\mathrm{s},m}+k_{\mathrm{g}}\Delta _{\mathrm{r},m} \right)}} \right|^2 \notag \\
    &=\frac{\alpha P \eta^4}{NM\sigma _{\mathrm{s}}^{2}}A^{\mathrm{(SA)}}\left( \mathbf{x}_{\mathrm{t}} \right) B^{\mathrm{(SA)}}\left( \mathbf{x}_{\mathrm{r}} \right). \label{con:sa-two-user-1a1}
\end{align}
To solve this problem, we first derive a solution for the PA placement on the Rx-SWAN.
More specifically, as the objective function in \eqref{obj:sa-two-user-1} is independent of $\mathbf{x}_{\mathrm{r}}$, and since $\mathbf{x}_{\mathrm{r}}$ and $\mathbf{x}_{\mathrm{t}}$ are separable in constraint \eqref{con:sa-two-user-1a1}, the optimal PA positions on the Rx-SWAN should drive the sensing SNR to reach the sensing threshold, i.e.,
\begin{align}
    \mathbf{x}_{\mathrm{r}}^{\star}=\mathrm{arg}\max \nolimits_{\mathbf{x}_{\mathrm{r}}\in \mathcal{F} _{\mathrm{r}}}B^{(\mathrm{SA)}}\left( \mathbf{x}_{\mathrm{r}} \right).
\end{align}
To this end, we adopt the coarse-to-fine strategy.
First, we select the waveguide containing the $x$-coordinate of the ST, whose index is denoted as $m^\star$.
Then, we place the PA for this segment at position $x_{\mathrm{t}, m^\star}=x_{\mathrm{s}}$ to minimize pathloss.
To identify the position of the remaining PAs, two steps are needed: \emph{1) coarse positioning to minimize path-loss}, and \emph{2) fine tuning to align phase shifts}.
First, we consider the PAs on the right-hand side of the $m^\star$-th segment.
Constrained by the minimal inter-spacing between the PAs, the first step is completed by 
\begin{align}
    \tilde{x}_{\mathrm{r}, m^\star+1} = \max \{x_{\mathrm{r}, m^\star}+\Delta_{\min}, x_{\mathrm{r}, m^\star+1}^{\mathrm{FD}} \}. \label{eq:coarse_path_loss_mitigation}
\end{align}
Upon obtaining the coarse position \eqref{eq:coarse_path_loss_mitigation}, the fine-tune step adds a deliberately designed position shift to ensure a constructive superposition on the signals, which can be obtained from
\begin{align}
   &F(\phi_{\mathrm{r},m^\star+1})\triangleq-k_{\mathrm{c}}(\left( \tilde{x}_{\mathrm{r},m^{\star}+1}^{}+\phi _{\mathrm{r},m^{\star}+1}^{}-x_{m^{\star}} \right) ^2+\Delta _{\mathrm{r},\mathrm{s}}^{2})^{1/2}\notag \\
   &-k_{\mathrm{g}}\left( x_{\mathrm{r},m^{\star}+1}^{\mathrm{FD}} \right. \left. -\tilde{x}_{\mathrm{r},m^{\star}+1}^{}-\phi _{\mathrm{r},m^{\star}+1}^{} \right) +k_{\mathrm{c}}(\left( x_{\mathrm{r},m^{\star}}^{\mathrm{FD}}-x_{m^{\star}} \right) ^2\notag \\
   &+\Delta _{\mathrm{r},\mathrm{s}}^{2})^{1/2}+k_{\mathrm{g}}\left( x_{\mathrm{r},m^{\star}}^{\mathrm{FD}}-x_{\mathrm{r},m^{\star}}^{} \right) =2\pi l
, \label{eq:phase_align_first}
\end{align}
where $l$ is determined by $l=\mathrm{Round}\{F\left( \phi _{\mathrm{r},m^{\star}+1}=0 \right) /(2\pi )\}$ and $\phi _{\mathrm{r},m^{\star}}^{}$ indicates the deliberate position shift to align the phase between the PAs.
A closed-form solution to \eqref{eq:phase_align_first} is given by
\begin{align}
\label{eq:phi_solution_general}
\phi _{\mathrm{r},m^{\star}+1}=\frac{I_1+\sqrt{I_{1}^{2}-\left( 1-n_{\mathrm{eff}}^{2} \right) I_2}}{1-n_{\mathrm{eff}}^{2}}-\tilde{x}_{\mathrm{r},m^{\star}+1},
\end{align}
where we need the following definitions:
\begin{align*}
I_1&=x_{\mathrm{s}}+n_{\mathrm{eff}}I_3, \qquad I_2=\Delta_{\mathrm{r,s}}^2 + x_{\mathrm{s}}^2-I_3^2, \\
I_3&=-\frac{2\pi l}{k_{\mathrm{c}}}+n_{\mathrm{eff}}(x_{m^{\star}}-x_{m^{\star}}^{\mathrm{FD}}-x_{m^{\star}+1}^{\mathrm{FD}})+\sqrt{x_{m^{\star}}^{2}+\Delta _{\mathrm{r},\mathrm{s}}^{2}}.
\end{align*}
Building on the above and letting $q\in \mathcal{Z}$ and $1 \le q\le M -m^\star$, the position of the $(m^\star+q)$-th PA is given by
\begin{align}
    x_{\mathrm{r}, m^\star+q} =\tilde{x}_{\mathrm{r}, m^\star+q}+\phi _{\mathrm{r},m^{\star}+q}.
\end{align}
Then, the next PA position indexed by $(m^\star+q)$ is obtained by a prerequisite coarse placement, detailed as follows:
\begin{align}
    \tilde{x}_{\mathrm{r}, m^\star+q} = \max \{x_{\mathrm{r}, m^\star+q-1}+\Delta_{\min}, x_{\mathrm{r}, m^\star+q}^{\mathrm{FD}} \}.
\end{align}
Subsequently, this coarse placement is refined by
\begin{align}
    x_{\mathrm{r}, m^\star+q} =\tilde{x}_{\mathrm{r}, m^\star+q}+\phi _{\mathrm{r},m^{\star}+q},
\end{align}
which is obtained by leveraging \eqref{eq:phase_align_first} via $x_{\mathrm{r},m^{\star}+q}^{\mathrm{FD}}\rightarrow x_{\mathrm{r},m^{\star}+q-1}^{\mathrm{FD}}$, $\tilde{x}_{\mathrm{r},m^{\star}+q}\rightarrow \tilde{x}_{\mathrm{r},m^{\star}+q-1}$.
Therefore, by performing the two steps iteratively, the remaining PAs on the right-hand side can be obtained.
For the PAs on the left-hand side, the coarse-to-fine policy is still applicable.
However, due to the asymmetric placement of the feed points, the coarse step is replaced by
\begin{align}
    \tilde{x}_{\mathrm{r},m^{\star}-1}=\min \{x_{\mathrm{r},m^{\star}}-\Delta _{\min},x_{\mathrm{r},m^{\star}}^{\mathrm{FD}}\}.
\end{align}
Additionally, $l=\mathrm{Round}\{F\left( \phi _{\mathrm{r},m^{\star}+1}=0 \right) /(2\pi )\}-1$ is used to obtain $l$.
\begin{algorithm}[t!]
    \small
    \caption{Element-wise Algorithm for Solving \eqref{pb:sa-two-user-1}}
    \label{alg:1}
    \begin{algorithmic}[1]
        \STATE{initialize the optimization variables}
        \REPEAT
            \FOR{$n \in  \{1,\dots,N\}$}
            \STATE{update $x_{\mathrm{t}, n}$ by increasing \eqref{obj:sa-two-user-1}, while checking the feasibility of \eqref{con:sa-two-user-1a} and \eqref{con:sa-two-user-1b} through one-dimensional search}
            \ENDFOR
        \UNTIL{The fractional decrease of the objective value \eqref{obj:sa-two-user-1} falls below a predefined threshold}
    \end{algorithmic}
\end{algorithm}
Thus, the final PA position can be computed via
\begin{align}
    x_{\mathrm{r}, m^\star-1} =\tilde{x}_{\mathrm{r}, m^\star-1}+\phi _{\mathrm{r},m^{\star}-1}.
\end{align}
By doing this repeatedly via variable substitutions $x_{\mathrm{r},m^{\star}-q}^{\mathrm{FD}}\rightarrow x_{\mathrm{r},m^{\star}+q+1}^{\mathrm{FD}}$ and $\tilde{x}_{\mathrm{r},m^{\star}-q}\rightarrow \tilde{x}_{\mathrm{r},m^{\star}-q+1}$ with $q\in\mathbb{Z}$ and $-m^\star \le q \le M-m^\star$, the PA positions on the left-hand side are determined.

Finally, for the optimization of the Tx-SWAN, i.e., $\mathbf{x}_{\mathrm{t}}$, we adopt an element-wise algorithm, as detailed in \textbf{Algorithm \ref{alg:1}}.
For computational complexity, since the PA positions on the Rx-SWAN side are derived in closed form, the overall complexity arises from the element-wise search on the Tx-SWAN side and is given by $\mathcal{O}(I_{\mathrm{iter}} Q N)$, where $I_{\mathrm{iter}}$ and $Q$ denote the number of iterations and the number of search grids per segment, respectively.

\subsection{SM-Based ISAC} \label{eq:sm_based_isac_single_cu}
To determine the sensing-rate Pareto front under the SM protocol, the rate optimization problem under the sensing constraint needs to be solved, which is expressed as follows: 
\begin{problem}\label{pb:sm-two-user-1} 
  \begin{alignat}{2}
  \underset{\mathbf{x}_{\mathrm{t}},\mathbf{x}_{\mathrm{r}},\mathbf{w}_{\mathrm{t}}, \mathbf{w}_{\mathrm{r}}}{\max} &\quad \log _2\left( 1+\gamma _{\mathrm{c}}^{\left( \mathrm{SS} \right)} \right) 
 \label{obj:sm-two-user-1}
    \\ \mathrm{s.t.} &\quad  \gamma _{\mathrm{s}}^{\left( \mathrm{SS} \right)}\ge \Gamma _{\mathrm{sen}}  \label{con:ss-two-user-1a}\\
    & \quad \|\mathbf{w}_{\mathrm{t}}\|=\|\mathbf{w}_{\mathrm{r}}\|=1 \label{con:sm-two-user-1b}
    \\  &\quad \mathbf{x}_{\mathrm{t}} \in \mathcal{F}_{\mathrm{t}}~\text{and}~\mathbf{x}_{\mathrm{r}} \in \mathcal{F}_{\mathrm{r}}.\label{con:sm-two-user-1c}
   \end{alignat}
\end{problem} 
where the additional unit-power constraints \eqref{con:sm-two-user-1b} specify the total power of the beamforming and combining vectors. 
Similar to the SA case, we first jointly design $\mathbf{x}_{\mathrm{r}}$ and $\mathbf{w}_{\mathrm{r}}$ on the Rx-SWAN to maximize the sensing SNR for any given $x_{\mathrm{t}}$ and $\mathbf{w}_{\mathrm{t}}$.
By doing so, the feasible solution sets for $x_{\mathrm{t}}$ and $\mathbf{w}_{\mathrm{t}}$ can be enlarged, thus contributing to the rate maximization.
For the SM protocol, the echo signal SNR can be written as follows: 
\begin{align}
    &\gamma _{\mathrm{s}}^{\left( \mathrm{SM} \right)}\left( \mathbf{x}_{\mathrm{t}},\mathbf{x}_{\mathrm{r}}, \mathbf{w}_{\mathrm{t}}, \mathbf{w}_{\mathrm{r}}\right)=\frac{\alpha P \eta^4}{\sigma _{\mathrm{s}}^{2}}A^{\mathrm{(SM)}}\left( \mathbf{x}_{\mathrm{t}}, \mathbf{w}_{\mathrm{t}} \right) B^{\mathrm{(SM)}}\left( \mathbf{x}_{\mathrm{r}}, \mathbf{w}_{\mathrm{r}}\right).  \notag
\end{align}
In the above equation, we need the following definitions:
\begin{align}
    A^{\mathrm{(SM)}}\left( \mathbf{x}_{\mathrm{t}}, \mathbf{w}_{\mathrm{t}} \right) = \left| \sum_{n=1}^N{\frac{w_{\mathrm{t},n}}{r_{\mathrm{s},n}}\mathrm{e}^{-\mathrm{j}\left( k_{\mathrm{c}}r_{\mathrm{s},n}+k_{\mathrm{g}}\Delta _{\mathrm{t},n}+\phi _{\mathrm{t},n} \right)}} \right|^2, \notag \\
    B^{\mathrm{(SM)}}\left( \mathbf{x}_{\mathrm{r}}, \mathbf{w}_{\mathrm{r}}\right)=\left| \sum_{m=1}^M{\frac{ w_{\mathrm{r},m}}{d_{\mathrm{s},m}}\mathrm{e}^{-\mathrm{j}\left( k_{\mathrm{c}}d_{\mathrm{s},m}+k_{\mathrm{g}}\Delta _{\mathrm{r},m}+\phi _{\mathrm{r},m} \right)}} \right|^2, \notag
\end{align}
where the transmit beamforming and combining vectors are defined as $\mathbf{w}_{\mathrm{t}}=\left[ w_{\mathrm{t},1}\mathrm{e}^{-\mathrm{j}\phi _{\mathrm{t},1}},....,w_{\mathrm{t}, N}\mathrm{e}^{-\mathrm{j}\phi _{\mathrm{t},N}} \right] ^{\mathrm{T}}$ and $\mathbf{w}_{\mathrm{r}}=\left[ w_{\mathrm{r},1}\mathrm{e}^{-\mathrm{j}\phi _{\mathrm{r},1}},....,w_{\mathrm{r},M}\mathrm{e}^{-\mathrm{j}\phi _{\mathrm{r},M}} \right] ^{\mathrm{T}}$.
Therefore, to maximize $B^{\mathrm{(SM)}}\left( \mathbf{x}_{\mathrm{r}}, \mathbf{w}_{\mathrm{r}}\right)$, the combining vector can be designed using MRC. 
As such, the entries of $\mathbf{w}_{\mathrm{r}}^{\mathrm{MRC}}$ are given by
\begin{align}
    \left[ \mathbf{w}_{\mathrm{r}}^{\mathrm{MRC}} \right] _m=\frac{1/r_{\mathrm{s},m}\mathrm{e}^{\mathrm{j}\left( k_{\mathrm{c}}d_{\mathrm{s},m}+k_{\mathrm{g}}\Delta _{\mathrm{r},m} \right)}}{\sum_{m=1}^M{\left( 1/r_{\mathrm{s},m}^{2} \right)}}, ~\forall m \in \mathcal{M}.
\end{align}
Thus, for any $\mathbf{x}_{\mathrm{r}}$, we have
\begin{align}
    B^{\mathrm{(SM)}}\left( \mathbf{x}_{\mathrm{r}},\mathbf{w}_{\mathrm{r}}^{\mathrm{MRC}} \right) =\sum_{m=1}^M{\frac{1}{\left( {x}_{\mathrm{s},m}^{}-x_{\mathrm{s}}^{} \right) ^2+\Delta _{\mathrm{r},s}^{2}}}. \label{eq:B_SM}
\end{align}
According to \eqref{eq:B_SM}, the phase alignment is completed by MRC.
Thus, the remaining step is to minimize the path loss, which can be achieved by applying the coarse antenna placement derived in the previous section to greedily reduce the path loss, as the phase alignment among PAs is automatically achieved by MRC.
For the Tx-SWAN side, the optimal beamforming vector $\mathbf{w}_{\mathrm{t}}^\star$ for any given $\mathbf{x}_{\mathrm{t}}$ can be determined via a subspace method, as detailed in the following lemma:
\begin{lemma} \label{lemma:subspace}
    \normalfont For given PA positions, the optimal solution to the beamforming design is obtained as 
    \begin{align}
        \mathbf{w}_{\mathrm{t}}^{\star}=\begin{cases}
	\sqrt{P}\mathbf{h}_{\mathrm{c}}/\left\| \mathbf{h}_{\mathrm{c}} \right\| ,\\
	c_1\hat{\mathbf{h}}_{\mathrm{s}}+c_2\hat{\mathbf{h}}_{\mathrm{c}\bot \mathrm{s}},\\
\end{cases}\begin{array}{c}
	\mathrm{if}~P>\Gamma _P,\\
	\mathrm{otherwise},\\
\end{array}
    \end{align}
    where we used the following definitions:
    \begin{align}
        &\Gamma _P=\frac{\Gamma _{\mathrm{sen}}\left\| \mathbf{h}_{\mathrm{c}} \right\| ^2\sigma _{\mathrm{s}}^{2}}{\alpha \left\| \mathbf{f}_{\mathrm{s}}^{} \right\| ^2\left| \mathbf{h}_{\mathrm{s}}^{\mathrm{H}}\mathbf{h}_{\mathrm{c}} \right|^2}, \quad \hat{\mathbf{h}}_{\mathrm{s}}=\mathbf{h}_{\mathrm{s}}/\left\| \mathbf{h}_{\mathrm{s}} \right\|, \notag \\
        &\hat{\mathbf{h}}_{\mathrm{c}\bot \mathrm{s}}=\frac{\mathbf{h}_{\mathrm{c}}-\left( \hat{\mathbf{h}}_{\mathrm{s}}^{\mathrm{H}}\mathbf{h}_{\mathrm{c}} \right) \hat{\mathbf{h}}_{\mathrm{s}}}{\left\| \mathbf{h}_{\mathrm{c}}-\left( \hat{\mathbf{h}}_{\mathrm{s}}^{\mathrm{H}}\mathbf{h}_{\mathrm{c}} \right) \hat{\mathbf{h}}_{\mathrm{s}} \right\|},~c_1=\sqrt{\frac{\Gamma _{\mathrm{sen}}\sigma _{\mathrm{s}}^{2}}{\alpha \left\| \mathbf{f}_{\mathrm{s}}^{} \right\| ^2\left\| \mathbf{h}_{\mathrm{s}}^{} \right\| ^2}}\frac{\hat{\mathbf{h}}_{\mathrm{s}}^{\mathrm{H}}\mathbf{h}_{\mathrm{c}}}{\left| \hat{\mathbf{h}}_{\mathrm{s}}^{\mathrm{H}}\mathbf{h}_{\mathrm{c}} \right|}, \notag \\
        &c_2=\sqrt{P-\frac{\Gamma _{\mathrm{sen}}\sigma _{\mathrm{s}}^{2}}{\left\| \mathbf{f}_{\mathrm{s}}^{} \right\| ^2\left\| \mathbf{h}_{\mathrm{s}}^{} \right\| ^2}}\frac{\mathbf{h}_{\mathrm{c}\bot \mathrm{s}}^{\mathrm{H}}\mathbf{h}_{\mathrm{c}}}{\left| \mathbf{h}_{\mathrm{c}\bot \mathrm{s}}^{\mathrm{H}}\mathbf{h}_{\mathrm{c}} \right|}. \notag 
    \end{align}
\end{lemma}
\begin{IEEEproof}
According to \cite{liu2022cramer}, we consider both inactive and active sensing SNR conditions.
For the former case, the optimal solution can be derived under MRT.
For the latter case, the optimal solution must lie within the subspace spanned by $\hat{\mathbf{h}}_{\mathrm{s}}$ and $\hat{\mathbf{h}}_{\mathrm{c}\bot \mathrm{s}}$ through proof by contradiction.
Then we plug a linear combination of these vectors back into the original problem and determine the complex-valued weights, completing this proof.
\end{IEEEproof}
Therefore, the resulting problem depends solely on the PA positions and can be solved using the element-wise approach presented in \textbf{Algorithm~\ref{alg:1}}, with the optimization objective replaced by \eqref{obj:sm-two-user-1}.
As for the computational complexity, since the PA positions on the Rx-SWAN side as well as the transmit and combining beamforming are given in closed form, the overall complexity is determined by the element-wise search on the Tx-SWAN side and is given by $\mathcal{O}(I_{\mathrm{iter}} Q N)$.

\section{SWAN-Assisted ISAC for the Multi-CU Case} \label{sect:pinch_multiplexing}
In this section, we consider the multiple CU case, where CUs are scheduled in a TDMA manner, and a single ST is detected.
To minimize the optimization overhead and delays caused by PA movement, the pinch multiplexing scheme is integrated with TDMA, i.e., pinch position optimization is performed only once across all time slots \cite{jiang2025pinching}.
Conversely, power allocation is individually optimized for each time slot to enhance the system’s sum rate.

\subsection{SS-Based ISAC} \label{sect:ps_for_multi_user}
For the SS protocol, define the power allocated to different time slots $\mathbf{p} \triangleq [P_1, ..., P_{K}]^{\mathrm{T}}$.
Collecting the optimization variables in set $\mathcal{V}_{\mathrm{SS}}\triangleq\{\mathbf{p}, x_{\mathrm{t},n}, x_{\mathrm{r},m}, m, n\}$, the sum-rate maximization problem under a sensing constraint can be formulated as follows:
\begin{problem}\label{pb:ss-multi-user-1} 
  \begin{alignat}{2}
    \underset{\mathcal{V}_{\mathrm{SS}}}{\max} &\quad \frac{1}{K}\sum\nolimits_{k=1}^K{R_{\mathrm{c},k}^{(\mathrm{SS)}}} \label{obj:ss-multi-user-1}
    \\\mathrm{s.t.} &\quad   \sum\nolimits_{k=1}^K{P_k}\le P_{\max} \label{con:ss-multi-user-1a} \\
    & \quad \gamma _{\mathrm{s},k}^{\left( \mathrm{SS} \right)}\ge \Gamma _{\mathrm{sen}}, \forall k \in \mathcal{K} \label{con:ss-multi-user-1b} \\
    & \quad R_{\mathrm{c},k}^{(\mathrm{SS)}}\ge \Gamma _{\mathrm{com}}, \forall k \in \mathcal{K} \label{con:ss-multi-user-1c}\\
    & \quad x_{\mathrm{t}, n} \in \mathcal{F}_{n},~x_{\mathrm{r}, m} \in \mathcal{F}_{m},  \label{con:ss-multi-user-1e}
   \end{alignat}
\end{problem} 
where $\mathcal{F}_{n} \triangleq [L(n-1),  Ln]$ and $\mathcal{F}_{m} \triangleq [L(m-1),  Lm]$ are the feasible position sets.
Constraint \eqref{con:ss-multi-user-1a} limits the total transmit power below the overall power budge $P_{\mathrm{max}}$;
constraint \eqref{con:ss-multi-user-1b} specifies the quality of service (QoS) requirement on sensing SNR of the ST; 
constraint \eqref{con:ss-multi-user-1c} represents the QoS requirement on communication rate of the $k$-th CU.
It is evident that the original problem \eqref{pb:ss-multi-user-1} is not jointly convex with respect to (w.r.t.) all optimization variables.
To alleviate this issue, we first derive the optimal PA position on the Rx-SWAN.
Building on this, the closed-form optimal power allocation policy for problem \eqref{obj:ss-multi-user-1} is derived based on the KKT conditions.
Finally, an element-wise search is utilized to solve the resulting optimization problem w.r.t. the PA positions on the Rx-SWAN.
Following the above logic flow, the optimal PA position on the Rx-SWAN is provided in the following lemma.
\begin{lemma}\label{lemma2}
    \normalfont Given that the PA positions of the Tx-SWAN $\mathbf{x}_{\mathrm{t}}$ and the power allocation policy $\mathbf{P}$ are fixed, the optimal PA position on the Rx-SWAN, denoted as $x_{\mathrm{r}}^{\star}$, is given by
    $x_{\mathrm{r}}^{\star} = x_{\mathrm{s}}$.
\end{lemma}
\begin{IEEEproof}
Let $x_1$ and $x_2$ be two distinct points in the feasible $x$-coordinate region of the Rx-SWAN, i.e., $[0, D_x]$.
Besides, assume that $x_2$ is closer to the $x$-coordinate of ST, i.e., $|x_2 - x_{\mathrm{s}}| < |x_1 - x_{\mathrm{s}}|$.
Since the SNR of the echo signal is a monotonically decreasing function w.r.t. the distance between the PA on Rx-SWAN and the ST, $\tilde{\gamma}_{\mathrm{s},k}^{\left( \mathrm{SS} \right)}\left( x_2 \right) >\tilde{\gamma}_{\mathrm{s},k}^{\left( \mathrm{SS} \right)}\left( x_1 \right) $ holds, thereby yielding $\tilde{\Gamma}_{\mathrm{sen}}\left( x_2 \right) \le \tilde{\Gamma}_{\mathrm{sen}}\left( x_1 \right)$.
Consequently, the transmit power threshold for $P_k$ can be expressed as follows:
\begin{align}
    \underset{\Gamma _k\left( x_2 \right)}{\underbrace{\max \{\tilde{\Gamma}_{\mathrm{com}},\tilde{\Gamma}_{\mathrm{sen}}\left( x_2 \right) \}}} < \underset{\Gamma _k\left( x_1 \right)}{\underbrace{\max \{\tilde{\Gamma}_{\mathrm{com}},\tilde{\Gamma}_{\mathrm{sen}}\left( x_1 \right) \}}}. \label{eq:constraint_x1_x2}
\end{align}
It is apparent that inequality \eqref{eq:constraint_x1_x2} holds for $\forall k$.
In this case, the solution set of $\mathcal{P} \triangleq \{P_k\}_{k=1}^K$ obtained by using $x_2$ is larger than that obtained by using $x_1$, i.e., $\mathcal{S} (x_1)\subseteq \mathcal{S} (x_2)$, where $\mathcal{S} (x)$ denotes the feasible CU position set when the PA is placed on the Rx-SWAN at location $x$.
This is intuitive, since the solution space can be enlarged by applying a less stringent constraint.
Therefore, denoting the objective function as a function of $x_{\mathrm{r}}$, i.e., $f^{\mathrm{OBJ}}(\mathcal{P})$, we have 
\begin{align}
    \max _{\mathcal{P} \in \mathcal{S} \left( x_2 \right)}f^{\mathrm{OBJ}}\left( \mathcal{P} \right) \ge \max _{\mathcal{P} \in \mathcal{S} \left( x_1 \right)}f^{\mathrm{OBJ}}\left( \mathcal{P} \right),
\end{align}
which is attributed to the fact that the objective function is a component-wise strictly increasing function w.r.t. each of the elements contained in $\mathcal{P}$.
This completes the proof.
\end{IEEEproof}
Building on \textbf{Lemma \ref{lemma2}}, the segment selection and PA positioning can be optimized by 1)~identifying $x_{\mathrm{r}}^\star$ by discarding the boundaries of segments; 2)~selecting the segment containing $x_{\mathrm{r}}^\star$.
In the same manner, $m^\star$ and $x_{\mathrm{r}, m^\star}$ can be found.
On the other hand, the residual problem w.r.t $x_{\mathrm{t},n}$ and $n$ can be solved in a similar manner, i.e., temporarily discarding the boundaries of the segments.
However, since the optimal PA positions on the Tx-SWAN, $x_{\mathrm{t}}^\star$, are difficult, if not impossible, to obtain, the element-wise search in \textbf{Algorithm \ref{alg:1}} is applied here.
Once $x_{\mathrm{t}}^\star$ is obtained from the search process, the segment containing this coordinate will be selected and marked as $n^\star$, with the PA position given by $x_{\mathrm{t}, n^\star}=x_{\mathrm{t}}^\star$.
For each search grid, the remaining optimization problem w.r.t. power allocation can be recast as follows:
\begin{problem}\label{pb:ss-multi-user-2} 
  \begin{alignat}{2}
    \underset{\{P_k\}_{k=1}^K}{\max} &\quad \frac{1}{K}\sum\nolimits_{k=1}^K{\log_2(1+\gamma _{\mathrm{c},k}^{\left( \mathrm{SS} \right)})} \label{obj:ss-multi-user-2}
    \\ \mathrm{s.t.} &\quad \eqref{con:ss-multi-user-1a}, \eqref{con:ss-multi-user-1b},~\mathrm{and}~\eqref{con:ss-multi-user-1c}. \notag
   \end{alignat}
\end{problem} 
which is a convex optimization problem w.r.t. $\{P_k\}_{k=1}^K$.
Before proceeding, we simplify the notations first.
First, we define ${\gamma}_{\mathrm{s},k}^{\left( \mathrm{SS} \right)}=P_k\tilde{\gamma}_{\mathrm{s},k}^{\left( \mathrm{SS} \right)}$, and ${\gamma}_{\mathrm{c},k}^{\left( \mathrm{SS} \right)}=P_k\tilde{\gamma}_{\mathrm{c},k}^{\left( \mathrm{SS} \right)}$.
Then, constraint \eqref{con:ss-multi-user-1c} can be written as $P_k\ge \left( 2^{\Gamma _{\mathrm{com}}}-1 \right) /\tilde{\gamma}_{\mathrm{c},k}^{\left( \mathrm{SS} \right)}\triangleq \tilde{\Gamma}_{\mathrm{com}}$.
Similarly, the sensing QoS constraint in \eqref{con:ss-multi-user-1b} can also be recast as $P_k\ge \Gamma _{\mathrm{sen}}/\tilde{\gamma}_{\mathrm{s},k}^{\left( \mathrm{SS} \right)}\triangleq \tilde{\Gamma}_{\mathrm{sen}}$.
Therefore, letting ${\Gamma}_k\triangleq \max \{ \tilde{\Gamma}_{\mathrm{com}},\tilde{\Gamma}_{\mathrm{sen}} \} $, the Lagrangian function for problem \eqref{pb:ss-multi-user-2} can be written as
\begin{align}
    &\mathcal{L} ^{\left( \mathrm{SS} \right)}=-1/K\sum\nolimits_{k=1}^K{\log_2(1+\gamma _{\mathrm{c},k}^{\left( \mathrm{SS} \right)})}  \notag \\
    &+\lambda _0\left( \sum\nolimits_{k=1}^K{P_k}-P_{\max} \right)
    +\sum\nolimits_{k=1}^K{\lambda _k\left( \Gamma_k -P_k \right)},
\end{align}
where $\lambda_0 \ge 0$ and $\{\lambda_k\}_{k=1}^K \geq 0$ denote the Lagrangian multipliers.  
Therefore, the KKT conditions for the optimal transmit power $\forall k$ can be characterized as follows:
\begin{align}
    \begin{cases}
    	\tilde{\gamma}_{\mathrm{c},k}^{\left( \mathrm{SS} \right)}/\left( K \ln 2\left( 1+P_k\tilde{\gamma}_{\mathrm{c},k}^{\left( \mathrm{SS} \right)} \right) \right) -\lambda _0+\lambda _k =0, \\
    	\lambda _0\left( \sum\nolimits_{k=1}^K{P_k}-P_{\max} \right) =0,\forall k,\\
    	\lambda _k\left( \Gamma_k -P_k \right) =0,\forall k,\\
    	\lambda _0\ge 0, \lambda _k \ge 0 ,~\Gamma_k -P_k\le 0, \forall k, \\
        \sum\nolimits_{k=1}^K{P_k}-P_{\max}\le 0. \\
    \end{cases} \notag
\end{align}
Here, we omit some primal feasibility conditions for brevity.
Based on the KKT condition, the optimal power allocation policy can be derived according to the classic water-filling algorithm, thus yielding:
\begin{align}
    P_{k}^{\star}=\max \left\{ \Gamma_k ,W^{\star}-1/\tilde{\gamma}_{\mathrm{c},k}^{\left( \mathrm{SS} \right)} \right\}, \label{eq:water_filling}
\end{align}
where the optimal water level $W^\star$ can be obtained via bisection search on $\sum\nolimits_{k=1}^K{P^\star_k}-P_{\max}= 0$ in logarithmic time, i.e., $\mathcal{O}(\log(K))$.
Thus, the overall computational complexity is given by $\mathcal{O}(I_{\mathrm{iter}}(Q+\log(K)))$ with $Q$ being the number of the search grids along the entire waveguide.

\subsection{SA-Based ISAC} \label{sect:sa_based_isac_multi_cu}
By denoting the optimization variable set as $\mathcal{V}_{\mathrm{SA}} \triangleq \left\{ \mathbf{p}, \mathbf{x}_{\mathrm{r}},\mathbf{x}_{\mathrm{t}}\right\}$, the optimization problem can be formulated as 
\begin{problem}\label{pb:sa-multi-user-1} 
  \begin{alignat}{2}
    \underset{\mathcal{V}_{\mathrm{SA}}}{\max} &\quad \frac{1}{K}\sum\nolimits_{k=1}^K{R_{\mathrm{c},k}^{(\mathrm{SA)}}} \label{obj:sa-multi-user-1}
    \\ \mathrm{s.t.} &\quad   \sum\nolimits_{k=1}^K{P_k}\le P_{\max} \label{con:sa-multi-user-1a} \\
    & \quad \gamma _{\mathrm{s},k}^{\left( \mathrm{SA} \right)}\ge \Gamma _{\mathrm{sen}}, \forall k \in \mathcal{K}\label{con:sa-multi-user-1b} \\
    & \quad R_{\mathrm{c},k}^{(\mathrm{SA)}}\ge \Gamma _{\mathrm{com}}, \forall k \in \mathcal{K} \label{con:sa-multi-user-1c}\\
    & \quad \mathbf{x}_{\mathrm{t}} \in \mathcal{F}_{\mathrm{t}}~\text{and}~\mathbf{x}_{\mathrm{r}} \in \mathcal{F}_{\mathrm{r}}.  \label{con:sa-multi-user-1d}
   \end{alignat}
\end{problem} 
Similar to the SS case, we adopt a two-step method to solve problem \eqref{pb:sa-multi-user-1}.
In the first step, we identify the optimal PA positions on the Rx-SWAN, i.e., $\mathbf{x}_{\mathrm{r}}^\star$, such that the sensing SNR is maximized for a given $\{\mathbf{p}, \mathbf{x}_{\mathrm{t}}\}$.
To achieve this goal, the coarse-then-refine policy detailed in Section \ref{sect:sa_based_isac_single_cu} can be applied.
In the second step, we apply the element-wise algorithm in \textbf{Algorithm \ref{alg:1}} to identify $\mathbf{x}_{\mathrm{t}}$.
For each search step, the water-filling method, as explained in \eqref{eq:water_filling}, is utilized to evaluate the sum-rate (i.e., the objective function value) given the optimal power allocation policy.
Thus, the computational complexity can be derived as $\mathcal{O}(I_{\mathrm{iter}}(QN+\log(K)))$ with $Q$ denoting the number of search grids per segment.

\subsection{SM-Based ISAC}
For the SM protocol, we define the optimization variable set as $\mathcal{V}_{\mathrm{SM}} \triangleq \left\{ \mathbf{p}, \mathbf{x}_{\mathrm{r}},\mathbf{x}_{\mathrm{t}}, \mathbf{w}_{\mathrm{t}, 1}, ...,\mathbf{w}_{\mathrm{t}, K}  \mathbf{w}_{\mathrm{r}, 1}, ..., \mathbf{w}_{\mathrm{r}, K}\right\}$, where $\mathbf{w}_{\mathrm{t}, k}$ and $\mathbf{w}_{\mathrm{r}, k}$ are the beamforming and combining vectors for the $k$-th time slot.
Thus, the optimization problem can be formulated as 
\begin{problem}\label{pb:sm-multi-user-1} 
  \begin{alignat}{2}
    \underset{\mathcal{V}_{\mathrm{SM}}}{\max} &\quad \frac{1}{K}\sum\nolimits_{k=1}^K{R_{\mathrm{c},k}^{(\mathrm{SM)}}} \label{obj:sm-multi-user-1}
    \\ \mathrm{s.t.} &\quad   \sum\nolimits_{k=1}^K{P_k}\le P_{\max} \label{con:sm-multi-user-1a} \\
    & \quad \gamma _{\mathrm{s},k}^{\left( \mathrm{SM} \right)}\ge \Gamma _{\mathrm{sen}}, \forall k \in \mathcal{K} \label{con:sm-multi-user-1b} \\
    & \quad R_{\mathrm{c},k}^{(\mathrm{SM)}}\ge \Gamma _{\mathrm{com}}, \forall k \in \mathcal{K} \label{con:sm-multi-user-1c}\\
    & \quad \mathbf{x}_{\mathrm{t}} \in \mathcal{F}_{\mathrm{t}}~\text{and}~\mathbf{x}_{\mathrm{r}} \in \mathcal{F}_{\mathrm{r}}  \label{con:sm-multi-user-1d} \\
    & \quad \|\mathbf{w}_{\mathrm{r},k}\| = \|\mathbf{w}_{\mathrm{r},k}\| =1, \forall k \in \mathcal{K}. \label{con:sm-multi-user-1e}
   \end{alignat}
\end{problem} 
To solve \eqref{pb:sm-multi-user-1} with moderate complexity, we adopt an alternating optimization (AO) framework.
As a preliminary step, the optimal PA positions on the Rx-SWAN, denoted by $\mathbf{x}_{\mathrm{r}}^\star$, and the optimal combining vector, denoted by $\mathbf{w}_{\mathrm{r},k}^\star$, are obtained through coarse antenna placement and MRC, respectively, as detailed in Section~\ref{eq:sm_based_isac_single_cu}.
Note that, since only one ST is considered and the objective function is independent of the combining vector, we have $\mathbf{w}_{\mathrm{r}}^\star = \mathbf{w}_{\mathrm{r},1}^\star = \cdots = \mathbf{w}_{\mathrm{r},K}^\star$.
In this case, the echo SNR is maximized on the Rx-SWAN side for a fixed set of the remaining optimization variables.
With power allocation in the outer layer, it is difficult, if not impossible, to derive the optimal beamforming vectors for each time slot.
Thus, we present a heuristic yet low-complexity design.
In particular, we parameterize $\mathbf{w}_k$ for $\forall k$ with $\varepsilon_k$ by
\begin{align}
    \mathbf{w}_{\mathrm{t},k}(\varepsilon_k)=\sqrt{\left( 1-\varepsilon_k \right) P_k}\hat{\mathbf{h}}_{\mathrm{c}}^{}+\sqrt{\varepsilon_k P_k}\hat{\mathbf{h}}_{\mathrm{s}\bot \mathrm{c}}^{},
\end{align}
where $\hat{\mathbf{h}}_{\mathrm{s}\bot \mathrm{c}}^{}\triangleq ( \mathbf{h}_{\mathrm{s}}^{}-( \hat{\mathbf{h}}_{\mathrm{c}}^{\mathrm{H}}\mathbf{h}_{\mathrm{s}}^{} ) \mathbf{h}_{\mathrm{c}}^{} ) / \| \mathbf{h}_{\mathrm{s}}^{}-( \hat{\mathbf{h}}_{\mathrm{c}}^{\mathrm{H}}\mathbf{h}_{\mathrm{s}}^{}) \mathbf{h}_{\mathrm{c}}^{} \| $.
The intuition behind this is to reallocate power within the space spanned by $\hat{\mathbf{h}}_{\mathrm{c}}^{}$ and $\hat{\mathbf{h}}_{\mathrm{s}\bot \mathrm{c}}^{}$, since allocating power outside this space is meaningless for either maximizing the sum rate or satisfying the sensing constraint.
Accordingly, we introduce a coefficient vector $\boldsymbol{\varepsilon} = [\varepsilon_1, \ldots, \varepsilon_K]^{\mathrm{T}}$ for all time slots, where $\varepsilon_k \in [0,1]$.
By traversing each entry in $\boldsymbol{\varepsilon}$, the optimal beamforming vectors for all time slots can be determined.
However, jointly traversing $\boldsymbol{\varepsilon}$ and the PA positions leads to prohibitive computational complexity.
To this end, we employ the AO framework as follows.
In the first step, we fix $\boldsymbol{\varepsilon}$ and solve the residual problem using the element-wise algorithm described in Section~\ref{sect:sa_based_isac_multi_cu}.
In the second step, we fix $\left\{ \mathbf{p}, \mathbf{x}_{\mathrm{r}}, \mathbf{x}_{\mathrm{t}} \right\}$ and optimize the entries of $\boldsymbol{\varepsilon}$ sequentially via one-dimensional search.
These two steps are alternated until the fractional decrease in the objective value of \eqref{obj:sm-multi-user-1} falls below a predefined threshold.
The resulting computational complexity is $\mathcal{O}(I_{\mathrm{iter}}(\log{K}Q_1N + KQ_2))$, where $Q_1$ and $Q_2$ denote the number of search grids per segment and the number of search grids of an interval $[0,1]$, $\forall k$, respectively.

\section{Numerical Results}\label{sect:results}
In this section, we present numerical results to verify the correctness of the derivations and the effectiveness of the proposed algorithms.
The following parameters are utilized throughout the simulations unless otherwise specified.
The carrier frequency is set to $f=28~\mathrm{GHz}$ and the effective refractive index is set to $n_{\mathrm{eff}}=1.4$.
The in-waveguide loss is set to $\kappa=0.08~\mathrm{dB/m}$.
The transmit power is set to $P =20~\mathrm{dBm}$ and the noise power is set to $\sigma_{\mathrm{s}}^2 = \sigma_{\mathrm{c}}^2 = -90~\mathrm{dBm}$.
Furthermore, the reflection coefficient is set to $\beta=1$.
Considering the topology of the SWAN-assisted ISAC system, the height of the waveguide is set to $D=3~\mathrm{m}$ with a minimum inter-PA spacing of $\Delta_{\min}=\lambda_{\mathrm{c}}/2$, while the $y$-coordinates of the Tx-SWAN and Rx-SWAN are set to $y_{\mathrm{r}}=-5~{\mathrm{m}}$ and $y_{\mathrm{t}}=+5~{\mathrm{m}}$, respectively.
The position of ST is fixed at $\mathbf{x}_{\mathrm{r}} = [10,0~\mathrm{m}, -6~\mathrm{m}, 0~\mathrm{m}]$.
For the hyperparameters, we set the search resolution to $\Delta_x=10^{-2}~{\mathrm{m}}$ for the element-wise method and $\Delta_{\varepsilon}=0.1$ for the subspace searching.
The position of the CUs is uniformly sampled within a two-dimensional area $D_x \times D_y=20~\mathrm{m} \times 20~\mathrm{m}$.

\subsection{Sensing Performance Gain of SWAN}
\begin{figure}
    \centering
    \includegraphics[width=0.75\linewidth]{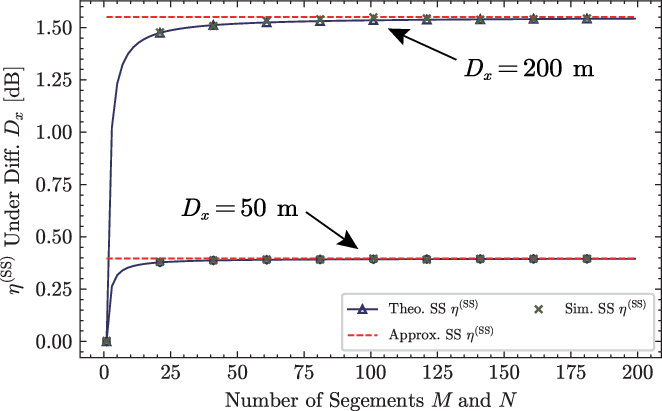}
    \caption{Illustration of SWAN's gain for the SS protocol and side lengths of $D_x=50~\mathrm{m}$ and $D_x=200~\mathrm{m}$ are considered, respectively.}
    \label{fig:1}
     \vspace{-1.5em}
\end{figure}
Fig. \ref{fig:1} depicts the performance gain of SWAN over conventional PASS transceivers for the SS protocol and the different lengths of the waveguides, i.e., $D_x=50~\mathrm{m}$ and $D_x=200~\mathrm{m}$.
As a reminder, this gain is evaluated by the ratio of sensing SNRs achieved by SWAN and conventional PASS.
For a fixed side length $D_x$, we see that as the number of segments increases, the gain over conventional PASS increases correspondingly and finally converges to the asymptotic results for large $M$ and $N$.
The reason is that, as the number of segments increases, the length of each segment decreases for a given sidelength $D_x$.
At the same time, the in-waveguide loss decreases, finally forming a loss-free continuous waveguide as $M,N\rightarrow\infty$.
Furthermore, by increasing the side length of the waveguide, SWAN's gain over conventional PASS can be further enhanced, since in this case, conventional PASS experiences more in-waveguide loss.

\begin{figure}
    \centering
    \includegraphics[width=0.75\linewidth]{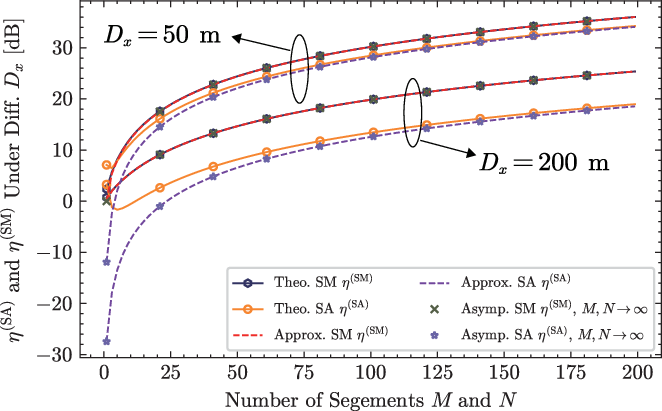}
    \caption{Illustration of SWAN's gain for the SA and SM protocols. Side lengths of  $D_x=50~\mathrm{m}$ and $D_x=200~\mathrm{m}$ are considered, respectively.}
    \label{fig:2}
     \vspace{-1.5em}
\end{figure}
Fig. \ref{fig:2} shows the performance gain of SWAN, i.e., sensing SNR ratios, for the SA and SM protocols, respectively.
The side length is varied to further investigate the impact of $D_x$.
For a given side-length configuration, the following observations are made: 1) For the SM case, the gain achieved by SWAN increases as the number of segments increases; 2) For the SA case, the gain initially decreases and subsequently increases as $M$ and $N$ become larger.
This suggests that the number of segments should be carefully chosen to maximize the benefits of SWANs.
Although the gain increases substantially with $M$ and $N$, it does not increase indefinitely due to manufacturing limitations and mutual coupling effects between antennas.
For large values of $M$ and $N$, the derived approximate solution yields results that closely match the theoretical values.
Furthermore, increasing the side length $D_x$ reduces the sensing SNR, since power dissipates into segments farther from the ST's position.

\subsection{Pareto Front of SWAN-Enabled ISAC}
\begin{figure}
    \centering
    \includegraphics[width=0.75\linewidth]{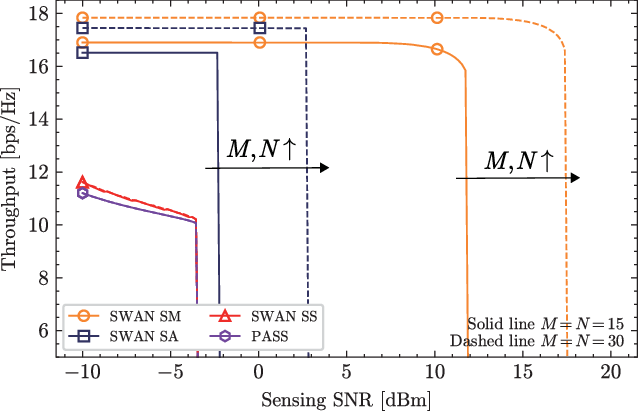}
    \caption{Illustration of the Pareto Front of the sensing and communication functionalities for the SS, SA, and SM protocols for $M=N=15$ and $M=N=30$ segments.}
    \label{fig:4}
    \vspace{-1.3em}
\end{figure}
\begin{figure}
    \centering
    \includegraphics[width=0.75\linewidth]{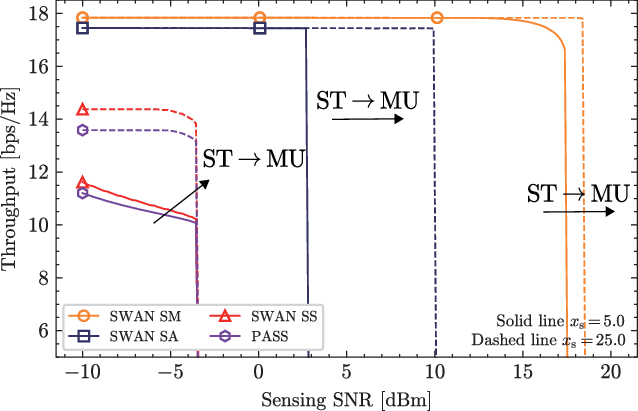}
    \caption{Illustration of the Pareto Front of the sensing and communication functionalities for the SS, SA, and SM protocols for ST $x$-coordinates $x_{\rm s}=5.0~\mathrm{m}$ and $x_{\rm s}=25.0~\mathrm{m}$, respectively. ``ST$\rightarrow$CU" indicates that the position of the ST is approaching that of the CU.}
    \label{fig:5}
     \vspace{-1.7em}
\end{figure}
Fig. \ref{fig:4} illustrates the Pareto front of sensing and communication for SWANs.
Fig. \ref{fig:4} is obtained by considering different sensing thresholds $\Gamma_{\mathrm{sen}}$.
In this figure, the coordinates of the CU and the ST are fixed at $\mathbf{r}_{\mathrm{c}}=[30~\mathrm{m}, 0~\mathrm{m}, 0~\mathrm{m}]$ and $\mathbf{r}_{\mathrm{s}}$
The figure shows that increasing the sensing SNR reduces the throughput, as pinching beamforming has to prioritize the sensing performance more.
A trade-off arises from the inherent competition between the communication and sensing functionalities.
With the highest degree of optimization (DoF), SM achieves the best performance.
SA exploits multiple segments and achieves a better performance than SS.
Furthermore, the reduction in in-waveguide loss achieved through segmentation results in an improved performance compared to conventional PASS.
By increasing the number of segments, the performance of the three protocols improves, except for the SS case, since this gain cannot be fully realized with a single segment.
This limitation arises because the in-waveguide loss is already minimal for moderate values of $M$ and $N$, so further increasing them yields marginal benefits.

Fig.~\ref{fig:5} depicts the Pareto front of the considered ISAC system as the ST's $x$-coordinate $x_{\mathrm{s}}$ varies, with the number of segments fixed at $M = N = 30$.
More specifically, when the ST's $x$-coordinate increases from $x_{\mathrm{s}} = 5~\mathrm{m}$ to $x_{\mathrm{s}} = 25~\mathrm{m}$, its distance to the CU at $x_{\mathrm{c}} = 30~\mathrm{m}$ decreases accordingly.
We observe that the Pareto front expands as the ST approaches the CU. 
This improvement is achieved because the sensing and communication objectives are more closely aligned due to their proximity.

\subsection{Pinch Multiplexing Enabled ISAC with Multiple CUs}
\begin{figure}
    \centering
    \includegraphics[width=0.75\linewidth]{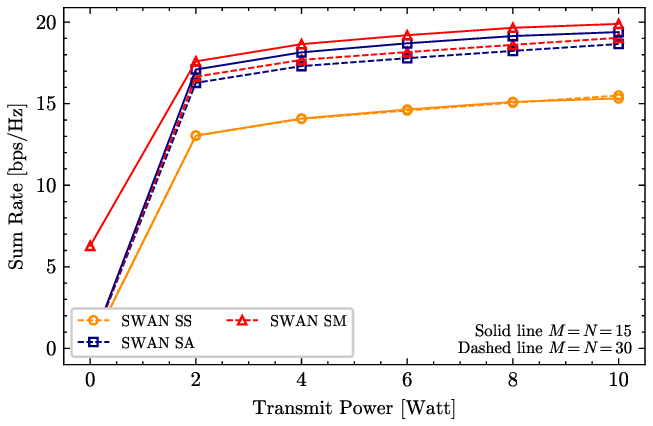}
    \caption{Illustration of sum rate versus the transmit power for different numbers of segments with $\Gamma_{\mathrm{sen}}=-50~\mathrm{dBm}$ and $K=3$.}
    \label{fig:6}
     \vspace{-1em}
\end{figure}
\begin{figure}
    \centering
    \includegraphics[width=0.75\linewidth]{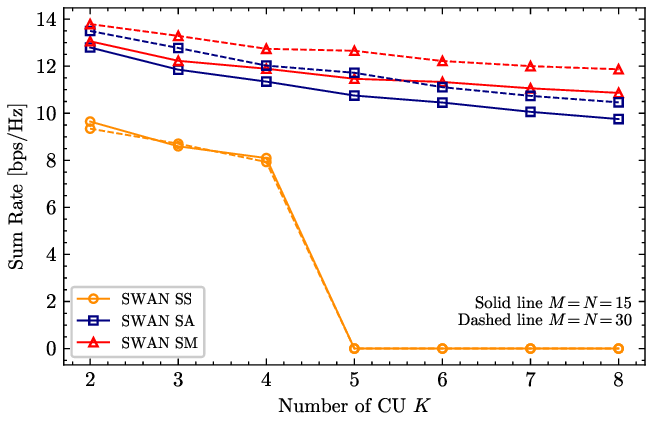}
    \caption{Illustration of sum rate versus the number of CUs for different numbers of segments with $\Gamma_{\mathrm{sen}}=-50~\mathrm{dBm}$ and $P_{\mathrm{max}}=0.1~\mathrm{W}$.}
    \label{fig:7}
     \vspace{-1.5em}
\end{figure}
Figs. \ref{fig:6} and \ref{fig:7} show results for the multi-CU scenarios enabled by pinching multiplexing, where the positions of the PAs are optimized once for all time slots at the beginning.
The sum rate is evaluated by $(1/K)\sum_{k=1}^K R_k$.
As shown in Fig. \ref{fig:6}, the sum rate for all protocols increases with increasing transmit power.
Furthermore, increasing the number of segments improves the performance of the SA and SM protocols.
However, the SS protocol's performance remains unchanged due to its segment-selection mechanism, i.e., only one segment is used.
Furthermore, when the transmit power is low, the sum rate becomes zero.
This is because, for low transmit powers, the sensing or communication requirements given in \eqref{con:ss-multi-user-1b} and \eqref{con:ss-multi-user-1c}, respectively, cannot be satisfied, which makes the problem infeasible.
Fig. \ref{fig:7} shows the sum rate versus the number of CUs.
As the number of CUs increases, the sum-rate decreases accordingly.
Additionally, the SM protocol demonstrates greater robustness to changes in the number of CUs, thanks to the higher DoF provided by full-digital beamforming.
For the SS protocol, a significant performance decline is observed when $K=5$.
This result arises because, for pinching multiplexing, the PA position is optimized only once for all time slots.
Within the SS protocol, only one segment is selected at the start of transmission.
As the number of CUs increases, a single PA position cannot satisfy the minimum communication threshold for all users, which renders the optimization problem infeasible.

\section{Conclusion}\label{sect:conclusions}
This paper investigated a SWAN-assisted ISAC system employing the SS, SA, and SM protocols.
First, a performance analysis validated the gain of SWAN over conventional PASS and demonstrated how this gain scales with the number of segments.
Subsequently, the one-user one-target case was analyzed to identify the Pareto front for sensing and communication for the three protocols.
Finally, for the general multi-CU scenario, pinch multiplexing-based TDMA was leveraged to schedule the CUs using element-wise algorithms to maximize the sum rate, subject to a sensing constraint.
Numerical results verified the derivations and the effectiveness of the proposed algorithms.

\bibliographystyle{IEEEtran}
\begin{spacing}{.96}
    \bibliography{refs}
\end{spacing}
\end{document}